\documentclass[%
superscriptaddress,
twocolumn,
 amsmath,amssymb,
 aps,
prb,
]{revtex4-1}

\usepackage{graphicx}% Include figure files
\usepackage{dcolumn}% Align table columns on decimal point
\usepackage{bm}% bold math
\usepackage{hyperref}% add hypertext capabilities
\renewcommand{\v}[1]{\ensuremath{\mathbf{#1}}} % for vectors
\newcommand{\gv}[1]{\ensuremath{\mbox{\boldmath$ #1 $}}}
\newcommand{\abs}[1]{\left| #1 \right|} % for absolute value
\newcommand{\pd}[2]{\frac{\partial #1}{\partial #2}}
\newcommand{\grad}[1]{\gv{\nabla} #1} % for gradient
\let\baraccent=\= % rename builtin command \= to \baraccent
\renewcommand{\=}[1]{\stackrel{#1}{=}} % for putting numbers above =

\newcommand{\nn}{\nonumber \\}

\begin{document}

\title{Role of generic scale invariance in a Mott transition from a U(1) spin-liquid insulator to a Landau Fermi-liquid metal}

\author{Jinho Yang}

\affiliation{Department of Physics, POSTECH, Pohang, Gyeongbuk 37673, Korea}

\author{Iksu Jang}

\affiliation{Department of Physics, POSTECH, Pohang, Gyeongbuk 37673, Korea}

\author{Jae-Ho Han}

\affiliation{Center for Theoretical Physics of Complex Systems, Institute for Basic Science (IBS), Daejeon 34051, Republic of Korea}

\author{Ki-Seok Kim}

\affiliation{Department of Physics, POSTECH, Pohang, Gyeongbuk 37673, Korea}

\date{\today}

\begin{abstract}
We investigate the role of generic scale invariance in a Mott transition from a U(1) spin-liquid insulator to a Landau Fermi-liquid metal, where there exist massless degrees of freedom in addition to quantum critical fluctuations. Here, the Mott quantum criticality is described by critical charge fluctuations, and additional gapless excitations are U(1) gauge-field fluctuations coupled to a spinon Fermi surface in the spin-liquid state, which turn out to play a central role in the Mott transition. An interesting feature of this problem is that the scaling dimension of effective leading local interactions between critical charge fluctuations differs from that of the coupling constant between U(1) gauge fields and matter-field fluctuations in the presence of a Fermi surface. As a result, there appear dangerously irrelevant operators, which can cause conceptual difficulty in the implementation of renormalization group (RG) transformations. Indeed, we find that the curvature term along the angular direction of the spinon Fermi surface is dangerously irrelevant at this spin-liquid Mott quantum criticality, responsible for divergence of the self-energy correction term in U(1) gauge-field fluctuations. Performing the RG analysis in the one-loop level based on the dimensional regularization method, we reveal that such extremely overdamped dynamics of U(1) gauge-field fluctuations, which originates from the emergent one-dimensional dynamics of spinons, does not cause any renormalization effects to the effective dynamics of both critical charge fluctuations and spinon excitations. However, it turns out that the coupling between U(1) gauge-field fluctuations and both matter-field excitations still persists at this Mott transition, which results in novel mean-field dynamics to explain the nature of the spin-liquid Mott quantum criticality. We discuss physical implications of effective one-dimensional spin dynamics and extremely overdamped gauge dynamics at the Mott quantum criticality.
\end{abstract}

\maketitle

\section{Introduction}

Although quasi-two-dimensional organic charge transfer salts such as $(BEDT-TTF)_{2}X$ have their complex structures given by molecular clusters, an effective Hubbard-type model on the triangular lattice system has been suggested to describe low-energy dynamics of these strongly correlated electrons at half filling \cite{Organic_Mott_Ins_Review}. In particular, $\kappa-(BEDT-TTF)_{2}Cu_{2}(CN)_{3}$ exhibited their paramagnetic Mott insulating behaviors at low temperatures, where NMR measurements did not show any characteristic features such as line broadening or spectrum shift \cite{Kappa_BEDT_NMR}. Moreover, heat capacity measurements confirmed the temperature-linear coefficient, implying the existence of a Fermi surface in the Mott insulating phase \cite{Kappa_BEDT_Specific_Heat}. On the other hand, thermal conductivity measurements suggested a gap-like behavior at lower temperatures, implying potential instability of the Fermi surface \cite{Kappa_BEDT_Thermal_Conductivity}. These experimental results seem to indicate that the paramagnetic Mott insulating phase may be identified with a spin liquid state. Furthermore, the Lieb-Schultz-Mattis (LSM) theorem \cite{LSM_Theorem} and its higher-dimensional generalizations by Oshikawa \cite{LSM_Oshikawa} and Hastings \cite{LSM_Hastings}, which states that a translation-invariant lattice model of the spin-$1/2$ system can not have a short-range entangled ground state preserving both spin and translation symmetries, implies that a symmetric gapped ground state has to be topologically ordered, in other words, a quantum spin liquid \cite{M_Cheng_LSM_Generalization}.

An interesting point in these quasi-two-dimensional organic charge transfer salts is that the Mott quantum criticality is universal regardless of the nature of the ground state \cite{Kanoda_MQCP_Transport}. For example, both $\kappa-(BEDT-TTF)_{2}Cu_{2}(CN)_{3}$ and $EtMe_{3}Sb[Pd(dmit)_{2}]_{2}$ are paramagnetic Mott insulators, suggesting spin liquid ground states, while $\kappa-(BEDT-TTF)_{2}Cu[N(CN)_{2}]Cl$ shows an antiferromagnetic Mott insulating ground state \cite{Organic_Mott_Ins_Review}. However, it turns out that the scaling behavior of the electrical resistivity in the vicinity of the Mott transition at finite temperatures shows the universal bifurcation behavior as a function of the scaled temperature, where the critical scaling exponent $\nu z$ for the energy scale is similar to each other within the range from $0.49$ to $0.68$ \cite{Kanoda_MQCP_Transport}. Here, $\nu$ is the correlation length critical exponent and $z$ is the dynamical critical exponent. $\nu z$ is the correlation energy-scale critical exponent. Considering that all these Mott transitions occur at finite temperatures even above the Neel temperature, we suspect that the universality of the Mott quantum criticality in these quasi-two-dimensional organic charge transfer salts may be involved with essentially the same quantum phase transition from a spin-liquid Mott insulator to a Landau Fermi-liquid metal. Recently, this universality has been investigated based on the dynamical mean-field theory (DMFT) framework \cite{DMFT_Mott_QCP}. Here, the Mott quantum criticality originates from the emergence of localized magnetic moments in the vicinity of the Mott transition, where their critical dynamics occurs only in the time direction, thus characterized by the infinite dynamical critical exponent. Interestingly, this DMFT framework could explain the universal bifurcation behavior of the electrical resistivity as a function of the scaled temperature, where the theoretical value of the energy-scale critical exponent $\nu z$ shows reasonable match with experiments.

In the present study, we investigate spin-liquid Mott quantum criticality for a Mott transition from a U(1) spin-liquid state to a Landau's Fermi-liquid phase based on the renormalization group (RG) analysis. An interesting feature of this problem is that there are two types of massless excitations near the quantum critical point. One massless excitations are critical charge fluctuations involved with this Mott transition, and the other gapless ones are U(1) gauge-field fluctuations coupled to a spinon Fermi surface in the spin-liquid state. This physical situation is in contrast to the conventional case, referred to as generic scale invariance \cite{Generic_Scale_Invariance}, where there exist massless degrees of freedom in addition to quantum critical fluctuations. This generic scale invariance gives rise to the fact that the scaling dimension of effective leading local interactions between critical charge fluctuations differs from that of the coupling constant between U(1) gauge fields and matter-field fluctuations in the presence of a Fermi surface \cite{Han_Cho_Kim_SLMQCP}. As a result, there appear dangerously irrelevant operators \cite{Dangerously_Irrelevant_OP}, which can cause conceptual difficulty in the implementation of RG transformations. Indeed, we find that the curvature term along the angular direction of the spinon Fermi surface is dangerously irrelevant at this spin-liquid Mott quantum criticality, responsible for divergence of the self-energy correction term in U(1) gauge-field fluctuations.

Performing the RG analysis in the one-loop level based on the technique of dimensional regularization, we reveal that such extremely overdamped dynamics of U(1) gauge-field fluctuations, which originates from the emergent one-dimensional dynamics of spinons, does not cause any renormalization effects to the effective dynamics of both critical charge fluctuations and spinon excitations. However, it turns out that the coupling between U(1) gauge-field fluctuations and both matter-field excitations still persists at this Mott transition, which results in novel mean-field dynamics to explain the nature of the spin-liquid Mott quantum criticality.

This paper is organized as follows. First, we introduce an effective field theory for spin liquid Mott quantum criticality in section \ref{EFT_SLMQCP}. Reviewing on the U(1) slave-rotor representation of the Hubbard Hamiltonian \cite{U1SR_Original}, we construct an effective field theory which consists of two sectors: Spin dynamics given by the ansatz of a U(1) spin-liquid ground state is described by spinons with a Fermi surface, and critical charge dynamics for the Mott transition is represented by ``relativistic" holons with its typical local interaction term, both of which are coupled to U(1) gauge-field fluctuations \cite{Han_Cho_Kim_SLMQCP}. Second, we perform the RG analysis of the one-loop level in section \ref{RG_EFT}. Our scaling analysis for this effective field theory confirms the existence of a dangerously irrelevant operator, here, the curvature term of the angular direction of the spinon Fermi surface, which causes the divergence of a Landau damping term for U(1) gauge-field fluctuations. Then, we introduce an effective renormalized action and its counter terms for the preparation of RG transformations, regarded to be a completely typical procedure. We calculate such counter terms in the one-loop level with the introduction of the divergent Landau damping term and obtain RG $\beta-$functions for coupling constants and Callan-Symanzik equations for correlation functions. Third, we summarize our main results and their physical implications in the concluding section \ref{Conclusion}.

Before going to the main body of this paper, we would like to point out the main difference between the present study and our recent investigation \cite{Han_Cho_Kim_SLMQCP}, performed by two of the authors. First of all, the divergent self-energy in gauge field which results from the dangerously irrelevant curvature term has been introduced into the RG analysis of essentially the same effective field theory. This extremely overdamped dynamics of U(1) gauge-field fluctuations does not cause any renormalization effects to the effective dynamics of both critical charge fluctuations and spinon excitations beyond the previous investigation. Although these two papers share the emergence of effective one-dimensional spin dynamics in a qualitative aspect, RG $\beta-$functions of coupling constants and critical exponents in correlation functions differ from each other as a result of the generic scale invariance.

\section{Effective field theory for spin-liquid Mott quantum criticality} \label{EFT_SLMQCP}

\subsection{Review on the U(1) slave-rotor representation of the Hubbard Hamiltonian}

We start our discussions, reviewing the U(1) slave-rotor representation of the Hubbard Hamiltonian, which describes a Mott transition from a U(1) spin-liquid insulator to a Landau's Fermi-liquid metal \cite{U1SR_Original}. Since this construction has been discussed in various previous studies \cite{U1SR_Kim,SU2SR_Kim}, one who is familiar to this parton construction may skip this review section.

We consider the partition function
\begin{eqnarray} && Z = \int D c_{i\sigma} \exp\Big[- \int_{0}^{\beta} d \tau \Big\{ \sum_{i} c_{i\sigma}^{\dagger} (\partial_{\tau} - \mu) c_{i\sigma} + H \Big\} \Big] , \nonumber \end{eqnarray}
where the Hubbard Hamiltonian is given by
\begin{eqnarray}
&& H = - t \sum_{i j}(c_{i \sigma}^\dagger c_{j \sigma} + H.c.) + U \sum_i n_{i \uparrow} n_{i \downarrow} .
\end{eqnarray}
Here, $c_{i \sigma}$ is an electron field with spin $\sigma = \uparrow, ~ \downarrow$ at site $i$ in the path integral representation, and $n_{i \sigma} = c_{i \sigma}^\dagger c_{i \sigma}$ is an electron-density field with spin $\sigma$. $\mu$ is the chemical potential to control the filling of electrons, here fit to be at half filling. $t$ is the hopping integral between nearest neighbor sites $i$ and $j$, and $U$ is the strength of on-site Hubbard interactions. The Einstein convention has been used for the spin summation in the kinetic energy term. $\beta$ is the inverse temperature in the partition function. We note that quasi-two-dimensional organic charge transfer salts form anisotropic triangular lattice structures, resulting in anisotropic hopping integrals \cite{anisotropy}.

%
%$t'/t = 0.8$ with a spin-liquid phase ($\kappa$-(BEDT-TTF)$_2$Cu$_2$(CN)$_3$) has anisotropic triangular lattice ($t'/t = 0.8$), but here we consider isotropic case with nearest neighbor hopping only, assuming strong frustration makes spin-liquid ground state.
%

Decomposing the interaction term into the charge and spin part as $n_{i \uparrow} n_{i \downarrow} = \frac{1}{2}(n_{i\uparrow}+n_{i\downarrow})^2 - \frac{1}{2}(n_{i\uparrow}-n_{i\downarrow})^2$, one may decompose spin part into the spin singlet and triplet sectors, where $U_{c}(1) \times U_{s}(1)$ global symmetry has been considered. Here, the subscripts $c$ and $s$ represent charge and spin, respectively. One can generalize this decomposition into the way to manifest $SU_{c}(2) \times SU_{s}(2)$ global symmetry \cite{SU2SR_Kim}. Since we are considering the U(1) spin-liquid ground state as an ansatz for the Mott insulating phase of $\kappa-(BEDT-TTF)_{2}Cu_{2}(CN)_{3}$, spin-triplet excitations are gapped, thus safely neglected in the present study. Performing the Hubbard-Stratonovich transformation for this spin singlet sector, we obtain
\begin{eqnarray}
Z &=& \int D c_{i\sigma} D \phi_{i} \exp\Big[- \int_{0}^{\beta} d \tau \Big\{ \sum_i c_{i \sigma}^{\dagger} (\partial_\tau -\mu + i \phi_i) c_{i \sigma} \nonumber \\ && - t \sum_{i j}(c_{i \sigma}^\dagger c_{j \sigma} + H.c.) + \frac{1}{U} \sum_i \phi_i^2 \Big\} \Big] , \label{HS}
\end{eqnarray}
where $\phi_i$ is an effective potential field.

To realize a spin-liquid state, we consider a parton construction referred to as U(1) slave-rotor representation \cite{U1SR_Original}: Separating the electron field into the bosonic charge part (holon) and the fermionic spin part (spinon) in the following way
\begin{eqnarray}
c_{i \sigma} = e^{-i \theta_i} f_{i \sigma} ,
\end{eqnarray}
we can reformulate the effective partition function Eq. (\ref{HS}) as
\begin{widetext}
\begin{eqnarray}
Z &=& \int D f_{i\sigma} D \theta_{i} D \phi_{i} \exp\Big[- \int_0^\beta d\tau \Big\{ \sum_i f_{i \sigma}^{\dagger} (\partial_\tau -\mu + i \phi_i) f_{i \sigma} - t \sum_{i j}(f_{i \sigma}^\dagger e^{i \theta_i} e^{-i \theta_j} f_{j \sigma} + H.c.) + \frac{1}{U} \sum_i (\partial_\tau \theta_i + \phi_i)^2 \Big\} . \nonumber \\
\end{eqnarray}
\end{widetext}
We note that the effective potential field has been shifted from $\phi_i$ to $\phi_i + \partial_\tau \theta_i$ in the last term \cite{U1SR_Kim,SU2SR_Kim}.

The final step for the mean-field theory analysis in the parton construction is to decompose the kinetic-energy term in the following way
\begin{widetext}
\begin{eqnarray}
Z &=& \int D f_{i\sigma} D \theta_{i} D \chi_{ij}^{f} D \chi_{ij}^{b} D \phi_{i} \exp\Big[- \int_0^\beta d\tau \Big\{ \sum_i f_{i \sigma}^{\dagger} (\partial_\tau -\mu + i \phi_i) f_{i \sigma} - t \sum_{i j}(f_{i \sigma}^\dagger \chi_{ij}^{f} f_{j \sigma} + H.c.) \nn && + \frac{1}{U} \sum_i (\partial_\tau \theta_i + \phi_i)^2 - t \sum_{i j}(e^{-i \theta_j} \chi_{ji}^{b} e^{i \theta_i} + H.c.) + t \sum_{i j} (\chi_{ij}^{f} \chi_{ji}^{b} + H.c.) \Big\} .
\end{eqnarray}
\end{widetext}
Here, $t \chi_{ij}^{f}$ and $t \chi_{ji}^{b}$ are effective hopping integrals of spinons and holons, respectively, where both spin and charge dynamics are renormalized by strong correlations in the spin-liquid state. For the mean-field theory analysis, it is conventional to replace $e^{i \theta_j}$ with $b_j$, where the uni-modular constraint $b^\dagger_i b_i = 1$ is introduced into the effective action by a Lagrange multiplier field in the path integral formulation. This Lagrange multiplier field plays the role of an effective mass term for holons, which controls their condensation when $U/t$ is tuned. The previous saddle-point analysis confirmed the existence of uniform mean-field solutions given by $\langle f_{i \sigma}^\dagger f_{j \sigma} \rangle = \chi^b$ and $\langle b_i b_j^\dagger \rangle = \chi^f$ in the vicinity of the holon condensation transition, which shows that a spin-liquid Mott insulating phase can be realized in an anisotropic Hubbard model on the triangular lattice system \cite{Lee_Lee_U1SL}.

The next inevitable task is to investigate the stability of this mean-field theory analysis for the Mott transition. Introducing low energy fluctuations of all kinds of order parameter fields, which result from all types of Hubbard-Stratonovich transformations, into the above effective lattice field theory, we obtain
%
%\begin{eqnarray}
%\langle f_{i \sigma} f^\dagger _{j \sigma} \rangle = -\chi^b e^{-i a_{i j}}, \quad \langle b_i b_j^\dagger \rangle = \chi^f e^{-i a_{i j}}
%\end{eqnarray}
%with $i \lambda_i = \lambda$ and $\phi_i = 0$ giving half filling conditions. Now the resulting effective action is given by,
%
\begin{widetext}
\begin{eqnarray}
Z &=& Z_{c} \int D f_{i\sigma} D b_{i} D a_{ij} D a_{i\tau} D \lambda_{i} \exp\Big[- \int_0^\beta d\tau \Big\{ \sum_i f_{i \sigma}^{\dagger} (\partial_\tau - \mu_{eff} + i a_{i\tau}) f_{i \sigma} - t \chi^f \sum_{i j}(f_{i \sigma}^\dagger e^{i a_{i j}} f_{j \sigma} + H.c.) \nonumber \\
&& + \frac{1}{U} \sum_i (- i b_{i}^{\dagger} \partial_{\tau} b_{i} + a_{i\tau})^{2} + m^{2} \sum_{i} (b_{i}^{\dagger} b_{i} - 1) - t \chi^b \sum_{ij} (b_j^\dagger e^{i a_{ji}}b_i + H.c.) + i \sum_{i} \lambda_{i} (b_{i}^{\dagger} b_{i} - 1) \Big\} \Big] .
\end{eqnarray}
\end{widetext}
Here, $Z_{c} = e^{- \beta L^{2} (2 z t \chi^{f} \chi^{b})}$ is the contribution from the condensation energy of effective hopping order-parameter fields, where $z$ is the coordination number and $L^{2}$ is the total number of lattice sites. $\mu_{eff}$ is an effective chemical potential to keep the half-filling condition of electrons. $m^{2} \propto U/t - (U/t)_{c}$ is an effective mass parameter of holons, given by the mean-field theory analysis with respect to the Lagrange multiplier field for the holon constraint. $a_{i\tau}$ is the low-energy fluctuation field of $\phi_{i}$ around its saddle-point value, which can be interpreted as the time component of the U(1) gauge field. $a_{ij}$ is that of $\chi_{ij}^{f}$ and $\chi_{ji}^{b}$ near their uniform mean-field values, which can be identified with the spatial component of the U(1) gauge field. $\lambda_{i}$ is that of the Lagrange multiplier field to impose the uni-modular constraint around its saddle-point value, which gives rise to mass fluctuations. This hardcore rotor constraint is sometimes softened to be $i \sum_{i} \lambda_{i} (b_{i}^{\dagger} b_{i} - 1) \rightarrow \lambda \sum_{i} (b_{i}^{\dagger} b_{i})^{2}$, expected not to change the nature of this Mott quantum criticality.

It is essential to notice that this effective field theory has U(1) gauge symmetry, given by
\begin{eqnarray}
f_{i \sigma} \rightarrow e^{i \alpha_i} f_{i \sigma}, \quad b_i \rightarrow e^{i \alpha_i} b_i, \quad a_{ij} \rightarrow a_{ij} - \alpha_i + \alpha_j ,
\end{eqnarray}
where the electron field $c_{i \sigma} = b_{i}^{\dagger} f_{i \sigma}$ is also gauge invariant as it should be.

Now, it is straightforward to take the continuum limit of this lattice field theory, given by
\begin{widetext}
\begin{eqnarray}
Z &=& Z_{c} \int D f_{\sigma} D b D \v{a} D a_{\tau} \exp\Big[- \int_0^\beta d\tau \int d^2 x \Big\{ f_{\sigma}^{\dagger} \Big(\partial_\tau - \mu_{eff} + i a_{\tau} - t\chi^f \grad^2 \Big) f_{\sigma} + i t \chi^f \v{a} \cdot \Big(f^\dagger_\sigma \big(\grad{f_\sigma}\big) - \big(\grad f^\dagger_\sigma\big) f_\sigma \Big) \nn && + t \chi^f \v{a}^2 f^\dagger_\sigma f_\sigma + \frac{1}{U} \big(- i b^{\dagger} \partial_{\tau} b + a_{\tau} \big)^{2} + m^{2} b^{\dagger} b + \frac{\lambda}{4} (b^{\dagger} b)^{2} + b^\dagger ( - t \chi^b \grad^2) b + i t \chi^b \v{a} \cdot \Big(b^\dagger \big(\grad{b}\big) - \big(\grad{b^\dagger}\big) b\Big) + t \chi^b \v{a}^2 b^\dagger b \nn && + \frac{1}{4 e^2}f_{\mu \nu} f_{\mu \nu} \Big\} \Big] , \label{EFT_UV}
\end{eqnarray}
\end{widetext}
where $f_{\mu \nu} = \partial_\mu a_\nu - \partial_\nu a_\mu$ is the field strength tensor for U(1) gauge-field fluctuations with $\mu, ~ \nu = 0, ~ 1, ~ 2$ for $d=2$. The gauge-field dynamics arises from quantum fluctuations of both spinons and holons.

In this study, we solve this effective field theory beyond the saddle-point analysis and investigate the nature of the spin-liquid Mott quantum criticality based on the RG transformation in the one-loop level.

\subsection{Effective field theory for spin-liquid Mott quantum criticality}

The Mott transition from the U(1) spin-liquid Mott insulator to the Landau's Fermi-liquid metal is expressed by the Higgs transition in the effective field theory Eq. (\ref{EFT_UV}) when the holon mass parameter $m^{2} \propto U/t - (U/t)_{c}$ is tuned. Then, the spinon Fermi surface of the U(1) spin liquid state evolves into a real Fermi surface of the Landau's Fermi-liquid phase, where the quasiparticle weight is given by the probability amplitude of holon condensates. In spite of this smooth connection, there exists essential difference in the dynamics of Fermi surface fluctuations. To describe the low-energy dynamics of fermion excitations near the Fermi surface, it is natural to expand the dispersion relation near the Fermi surface. Then, one can see that the dispersion of the angular direction along the Fermi surface is dimensionless for the quasiparticle dynamics in the Landau's Fermi-liquid state. This scaling property is important in the Shankar's RG analysis for the Landau's Fermi-liquid state \cite{LFL_RG}. On the other hand, it has been shown that the dispersion of the angular direction along the Fermi surface acquires its anomalous scaling dimension when there are gapless excitations coupled to the Fermi surface \cite{Anomalous_Scaling_FS,EFT_Large_N}. To deal with this anomalous Fermi-surface dynamics in the presence of critical fluctuations, Sung-Sik Lee proposed a patch construction of the Fermi surface, where the dispersion of the angular direction plays the role of a curvature term in the linear dispersion near the Fermi surface. Dividing the Fermi surface into patches, he investigated the coupling nature between Fermi-surface patches in the presence of U(1) gauge-field fluctuations. Interestingly, scattering of spinons between different patches turns out to be irrelevant at low energies due to the anomalous scaling dimension of the angular directional dispersion, which originates from coupling to massless U(1) gauge-field excitations. As a result, a double-patch Fermi-surface model has been proposed for quantum critical dynamics of Fermi-surface excitations \cite{Double_Patch_Construction_I,Double_Patch_Construction_II}, where critical bosonic excitations occur at the zero ordering wave vector.

\begin{figure}
\centering
\includegraphics[width=8cm]{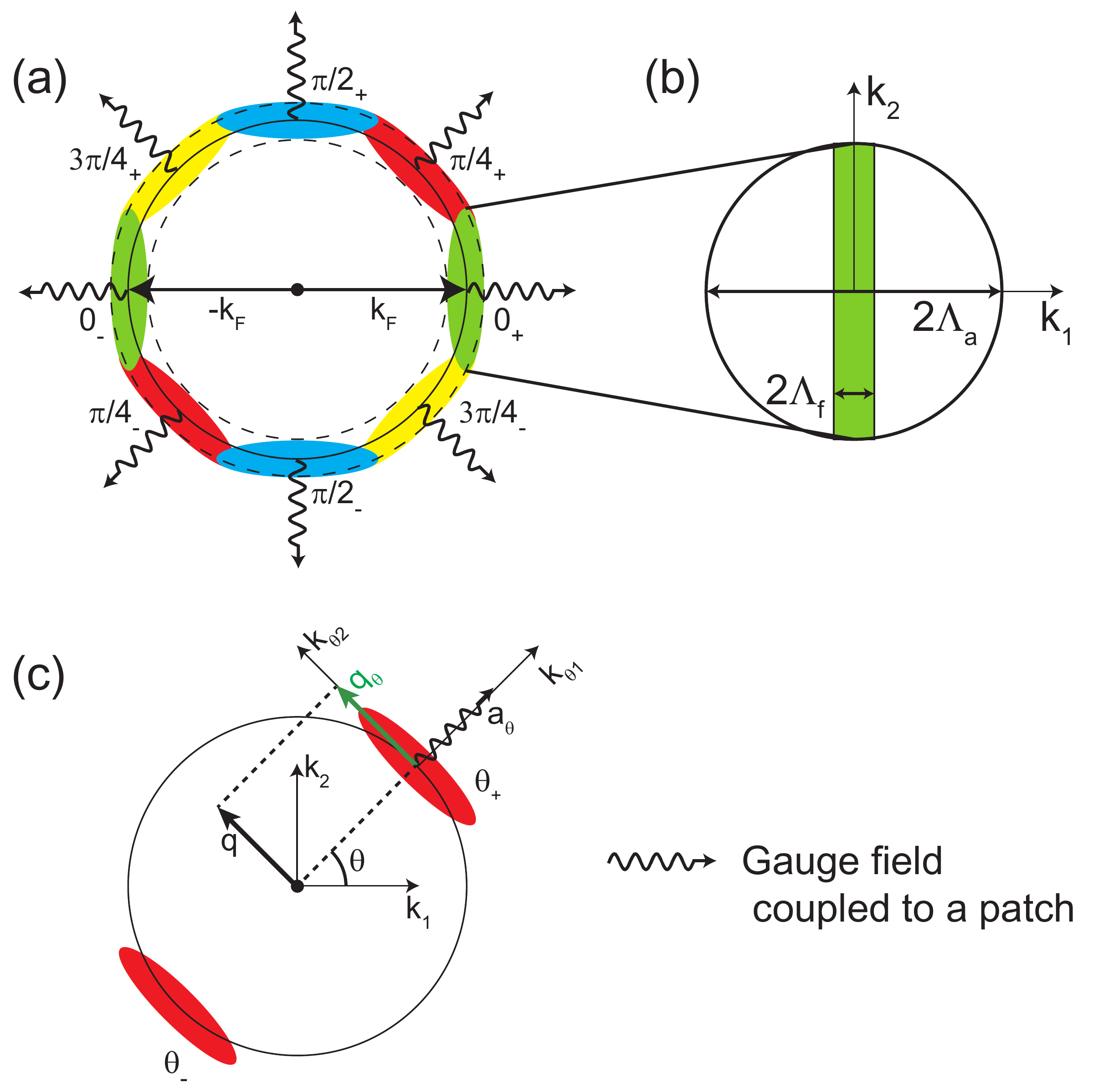}
\caption{The patch construction for a spinon Fermi surface. (a) Patches with a patch index $0 \leq \theta_s < \pi$. Here, $s = \pm$ denotes two partner patches in the minimal model of the double patch construction. (b) Two types of UV cutoffs near the spinon Fermi surface. $\Lambda_{f}$ and $\Lambda_{a}$ are spinon and gauge-field UV cutoffs, respectively, where $\Lambda_{a}$ is larger than $\Lambda_{f}$. $\Lambda_{a} \gg \Lambda_{f}$ is the origin of the property of patch decoupling \cite{Anomalous_Scaling_FS,EFT_Large_N}. (c) Coordinate transformation between the momentum of the global coordinate at the origin and that of the local patch coordinate at a Fermi-surface patch. Here, $q_{\theta}$ is the momentum of the U(1) gauge field $a_{\theta}$ in the local patch coordinate.}
\label{patchconstruction}
\end{figure}

Our situation is more complicated: In addition to U(1) gauge-field fluctuations coupled to the spinon Fermi surface, there are massless holon excitations involved with the Mott quantum criticality, also coupled to U(1) gauge-field fluctuations. In principle, we have to consider all directions of U(1) gauge fields because holon excitations interact with such U(1) gauge-field fluctuations with all momentum directions. We recall that only one direction of gauge-field excitations is enough to describe the U(1) spin-liquid state with a spinon Fermi surface, referred to as the double patch construction as discussed above, where charge fluctuations are gapped. In this respect it is not completely clear whether the patch-decoupling nature still survives or not. To check out the validity of the double patch construction in the present case, we revisit the patch construction, where the patch index is given by a subscript $\theta_s$. See Fig. \ref{patchconstruction} (a). Here, $\theta$ covers from $0$ to $\pi$ while $s = \pm$ indicates two partner patches for given $\theta$ in the double-patch minimal model construction, i.e., $\theta_{-} = \theta_{+} + \pi$. See Fig. \ref{patchconstruction} (c). Below, we do not sometimes show the subscript $s$ as the patch index for concise notation. For each Fermi-surface patch which is on angle $\theta$, we linearize the spinon dispersion as follows
\begin{eqnarray}
&& - i \omega - \mu + t \chi^f \v{k}^2 \rightarrow -i \omega + s ~ v_f k_{\theta,d-1} + t \chi^f k_{\theta,d}^2 ,
\end{eqnarray}
where $v_f = 2 t \chi^f k_F$ is the Fermi velocity and $0 \leq \theta < \pi$ is the patch index. We recall $s = \pm$ at given $\theta$ and $d = 2$. We point out that this local patch coordinate near the Fermi surface can be translated into the the momentum of the global coordinate at the origin as follows
\begin{eqnarray}
&& \left( \begin{array}{cc} \cos{\theta} & -\sin{\theta} \\ \sin{\theta} & \cos{\theta} \end{array} \right) \left( \begin{array}{c} k_{d-1} \\ k_{d} \end{array} \right) = \left( \begin{array}{c} k_{\theta, d-1} \\ k_{\theta, d} \end{array} \right) , \label{convert}
\end{eqnarray}
well shown in Fig. \ref{patchconstruction} (c), where $(k_{d-1}, k_{d})$ is the momentum of the global coordinate at the origin and $(k_{\theta, d-1}, k_{\theta, d})$ is that of the local patch coordinate at the Fermi surface. We point out that there are two types of UV cutoffs: $\Lambda_{f}$ and $\Lambda_{a}$ are spinon and gauge-field UV cutoffs, respectively, where $\Lambda_{a}$ is larger than $\Lambda_{f}$. $\Lambda_{a} \gg \Lambda_{f}$ is the origin of the property of patch decoupling \cite{Anomalous_Scaling_FS,EFT_Large_N}.

Based on these discussions, we obtain an effective field theory in the patch construction as follows
\begin{eqnarray}
S&=& S_f + S_b + S_a + S_{fa} + S_{ba} , \nonumber \\
S_f&=& \int_{k_\theta} f^\dagger_{\sigma \theta_s}(k_\theta) \left ( i k_0 + s ~ v_f k_{\theta,d-1} + t \chi^f k_{\theta,d}^2 \right ) f_{\sigma \theta_s}(k_\theta) , \nonumber \\
S_b&=& \int_k b^\dagger(k) \left (\frac{1}{U} k_0^2 + t \chi^b \v{k}^2 + m^{2} \right ) b(k) \nn &+& \frac{\lambda}{4} \int_{k,p,q} b^\dagger(k+q) b(k) b^\dagger(p-q) b(p) , \nonumber \\
S_a&=&\frac{1}{2} \int_q a_i(-q) ( q^2 \delta_{i j} - q_{i} q_{j} ) a_j(q) , \nonumber \\
&=& \frac{1}{2} \int_{q_\theta} a_{\theta}(-q_{\theta})(q_0^2 + \v{q_{\theta}}^2 ) a_{\theta}(q_{\theta}) , \nonumber \\
S_{fa}&=& -v_f e \int_{k_\theta,q_\theta} s ~ a_{\theta}(q_\theta) f^\dagger_{\sigma \theta_s}(k_\theta+q_\theta) f_{\sigma \theta_s}(k_\theta) \nonumber \\
&+& t \chi^f e^2 \int_{k_\theta,p_\theta,q_\theta} a_{\theta}(-p_\theta+q_\theta) a_{\theta}(p_\theta) f^\dagger_{\sigma \theta_s}(k_\theta+q_\theta) f_{\sigma \theta_s}(k_\theta), \nonumber \\
S_{ba} &=& - 2 t \chi^b e \int_{k,q} k_i a_i(q) b^\dagger(k+q) b(k) \nonumber \\
&+& t \chi^b e^2 \int_{k,p,q} a_i (-p+q) a_i (p) b^\dagger (k+q) b(k) .
\label{action0}
\end{eqnarray}
Here, we introduced $\int_k = \int d^{3}k/(2 \pi)^{3}$ for the integral expression, $q^2 = q_0^2 + \v{q}^2$ for the gauge-field dynamics in $S_a$, and the Einstein convention has been used for the $\theta$ index summation. First of all, we consider the Coulomb gauge $\v{q} \cdot \v{a}(q) = q_{i} a_{i}(q) = 0$ with $i = 1, ~ 2$, which gives rise to decoupling between potential and spatial gauge fluctuations. Then, the time component of the U(1) gauge field can be safely neglected in this expression because such potential fluctuations are gapped due to the presence of a Fermi surface, referred to as Debye screening \cite{Nagaosa_Lee_SL}. We emphasize that the fermion dynamics near the Fermi surface is described by the local patch coordinate $(k_{\theta, d-1}, k_{\theta, d})$ with $d = 2$ as shown in $S_f$ of Eq. (\ref{action0}) while the boson dynamics is expressed by the global coordinate $(k_{d-1}, k_{d})$ near the origin as shown in $S_b$ of Eq. (\ref{action0}). In this respect we show both coordinate expressions for gauge-field fluctuations as shown in $S_a$ of Eq. (\ref{action0}), coupled to both matter-field fluctuations, and these gauge-field actions are interchanged by the coordinate transformation Eq. (\ref{convert}).

We recall that spinons $f_{\sigma \theta_s}$ in a given patch couple to gauge-field excitations with a direction perpendicular to the patch, here denoted as $a_{\theta}$ in the same direction of $q_{\theta,d-1}$. See $S_{fa}$ of Eq. (\ref{action0}). As a result, such gauge-field fluctuations in all momentum directions will get quantum corrections from fermion excitations of the corresponding patch. On the other hand, holon excitations cause quantum corrections to gauge-field fluctuations of any momentum directions and vice versa. See $S_{ba}$ of Eq. (\ref{action0}). Again, this is the reason why the gauge-field dynamics is expressed by both local patch and global origin coordinates as shown in $S_{a}$ of Eq. (\ref{action0}). To investigate renormalization effects from $S_{ba}$ of Eq. (\ref{action0}), we resort to the global coordinate representation. On the other hand, when we examine gauge field-spinon quantum corrections, we consider the local patch coordinate. An essential point is that the decoupling nature of the patch construction still holds in the presence of critical charge fluctuations. Following Sung-Sik Lee's argument \cite{Anomalous_Scaling_FS,EFT_Large_N}, one can show that patch mixing does not occur for the $S_{f} + S_{a} + S_{fa}$ sector in the low-energy limit, which results from the anomalous scaling dimension of the Fermi-surface curvature term in the presence of gapless gauge-field fluctuations. Furthermore, we observe that it does not occur for the RG transformation of the $S_{b} + S_{a} + S_{ba}$ sector in the one-loop level, either, when the global coordinate representation is translated into the local patch representation. It is clear that the patch mixing cannot happen except for the partner patch of the double patch construction at least in the one-loop level RG analysis. Although we suspect that this patch-decoupling nature will be preserved beyond the one-loop RG analysis, we cannot give clear proof at present.

%
%Therefore, our strategy is to apply quantum corrections to the gauge field free part due to fermion using the second form, and then we convert the second form to the first form and evaluate quantum corrections between gauge field and holon. Note that one can always convert $a(q) \leftrightarrow a_{\theta}(q_\theta)$ using Eq. \ref{convert}.
%

To perform the RG transformation in this complex problem, we resort to the dimensional regularization technique. In particular, we extend the time dimension, keeping the spatial dimension being two \cite{Dimensional_Regularizarion_FS_I,Dimensional_Regularizarion_FS_II}. This regularization technique extends the codimension of the Fermi surface, preserving its dimension as one. Although this codimensional regularization method breaks the global symmetry of the original dimension, it does not cause UV-IR mixing, which means that the IR physics is not purely determined from its low-energy effective field theory but involved with its UV data, for example, the size of the Fermi surface, the UV cutoff, and etc.. The patch mixing is one source of the UV-IR mixing phenomena \cite{UV_IR_Mixing}, which does not arise in the one-loop RG analysis.
However, we claim that the Landau damping term has to be modified from the conventional case in the presence of a dangerously irrelevant operator, where the UV cutoff is introduced. This point will be discussed below in more details.
We combine two spinon fields of the partner patch in the double patch construction as follows \cite{Dimensional_Regularizarion_FS_I,Dimensional_Regularizarion_FS_II}
\begin{eqnarray}
\psi_{\sigma \theta} (k) = \left( \begin{array}{c} f_{\sigma \theta_+} (k_\theta) \\
 f^\dagger_{\sigma \theta_-} (k_\theta) \end{array} \right) .
\end{eqnarray}
Then, it is straightforward to take the codimensional regularization as \cite{Han_Cho_Kim_SLMQCP}
\begin{eqnarray}
S&=& S_f + S_b + S_a + S_{fa} + S_{ba}, \nonumber \\
S_f&=& \int_{k_\theta} \bar{\psi}_{\sigma \theta}(k_\theta) (i \boldsymbol{\Gamma} \cdot \v{K} + i \gamma_{d-1} \delta_{k_\theta}) \psi_{\sigma \theta}(k_\theta), \nonumber \\
S_b&=& \int_k b_a^\dagger(k) \left (\frac{1}{U} \v{K}^2 + t \chi^b \v{k}^2 \right ) b_a(k), \nonumber \\
&& + \frac{\lambda}{4 N} \int_{k,p,q} b_a^\dagger(k+q) b_a(k) b_{b}^\dagger(p-q) b_b(p) \nonumber \\
S_a&=&\frac{1}{2} \int_{q} a_i(-q)(\v{Q}^2 + \v{q}^2)\delta_{i j} a_j(q), \nonumber \\
S_{fa}&=& \frac{i v_f e}{\sqrt{N}} \int_{k_\theta,q_\theta} a_\theta(q_\theta) \bar{\psi}_{\sigma \theta}(k_\theta+q_\theta) \gamma_5 \gamma_{d-1} \psi_{\sigma \theta}(k_\theta) \nonumber \\
S_{ba} &=& -\frac{2t \chi^b e}{\sqrt{N}} \int_{k,q} k_i a_i (q) b_a^\dagger(k+q) b_a(k) \nonumber \\
&& + \frac{t \chi^b e^2}{N} \int_{k,p,q} a_i (-p+q) a_i (p) b_a^\dagger (k+q) b_a(k) . \nonumber \\
\label{action1}
\end{eqnarray}
Here, the frequency dimension given by $k_{0}$ and $q_{0}$ is extended to $d-2$ dimensions described by $\v{K} = (k_0,k_1,...,k_{d-2})$ and $\v{Q} = (q_0,q_1,...,q_{d-2})$, respectively. Accordingly, the Dirac gamma matrix $\gamma_{0}$ for the frequency sector is generalized as $\boldsymbol{\Gamma} = (\gamma_{0}, \gamma_{1}, ..., \gamma_{d-2})$, where a generalized form of $\gamma_{5} = \gamma_{0} \gamma_{1} \cdot\cdot\cdot \gamma_{d-2} \gamma_{d-1}$ has been also introduced into $S_{fa}$. Although the diamagnetic coupling term is not shown explicitly in $S_{fa}$, its role has to be taken into account in order to preserve the U(1) gauge symmetry in the perturbative RG analysis \cite{Gauge_Symmetry_Correlation_Functions}.
%
%Here, the Dirac gamma matrixes are given by $\gamma_0 = \sigma_y$, $\gamma_1 = \sigma_x$, and $\gamma_5 =  i \gamma_0 \gamma_1 = \sigma_z$, where $\sigma_{i}$ with $i = x, ~ y, ~, z$ are Pauli matrixes.
%
$\delta_{k_\theta} = v_f k_{\theta,d-1} + t \chi^f k^2_{\theta,d}$ is the linearized dispersion relation of spinons near the Fermi surface, where the anomalous scaling dimension will arise from the curvature term, clarified in the next section. To be consistent with the codimensional regularization in the frequency sector, the Dirac gamma matrix $\gamma_{1}$ is replaced with $\gamma_{d-1}$. $\bar{\psi}_{\sigma \theta}(k_\theta) = \psi^\dagger_{\sigma \theta}(k_\theta) \gamma_0$ is typically introduced in the spinor representation, and the integral expression is given by $\int_k = d^{d+1} k/(2 \pi)^{d+1}$ due to the dimensional regularization. Formally, we introduced the holon flavor index $a = 1, ..., N$ in $b_a(k)$ although this extension from $a = 1$ is not essential \cite{Han_Cho_Kim_SLMQCP}.

\section{Renormalization Group Analysis} \label{RG_EFT}

\subsection{Scaling analysis}

In this study we start our RG transformation from an intermediate fixed point which preserves the dispersion relation of holons. Considering the fact that we focus on the Mott quantum criticality, this choice for the intermediate UV fixed point looks natural \cite{Han_Cho_Kim_SLMQCP}. One may suggest to consider a Gaussian fixed point near the U(1) spin-liquid fixed point, which preserves the spinon dispersion relation. Since we start from the U(1) spin liquid state and focus on the evolution of the Fermi-surface dynamics, this intermediate fixed point also seems to be natural. However, technical difficulty arises in this case, where the fractional upper critical dimension $d_{c} = 5/2$ for the spinon-gauge field coupling constant does not allow us to extract out so called $\epsilon$ poles for the RG transformation of the holon sector in the dimensional regularization method \cite{RG_Textbook}. Here, $\epsilon = d_{c} - d$ is the distance from the upper critical dimension. Then, one has to find another regularization method for the consistent implementation of RG transformations. In this respect we consider the Gaussian fixed point to preserve the holon dynamics as our starting point.

To find the upper critical dimension of all types of interaction vertices, we consider the scaling transformation preserving the holon dispersion as
\begin{eqnarray}
\v{K} = \frac{\v{K}'}{s} , \quad \quad \vec{k} = \frac{\vec{k}'}{s} . \label{holonscale}
\end{eqnarray}
Then, we count the engineering dimension $\Delta_b$ of the holon field as
\begin{eqnarray}
b_a(k) = s^{\Delta_b} b'_a(k') , \quad \quad \Delta_b = \frac{d+3}{2}.
\end{eqnarray}
Based on this information, we obtain the scaling dimension of the coupling constant $\lambda$ as
\begin{eqnarray}
\lambda = s^{\Delta_\lambda} \lambda', \quad \quad \Delta_\lambda = d-3 ,
\end{eqnarray}
which shows that the upper critical dimension if this interaction vertex is $d_{c} = 3$ as expected.

In a similar way, one finds the engineering dimension of the U(1) gauge field $a_{i}(q)$ and the scaling dimension of the coupling constant $e$ as
\begin{eqnarray}
a_{i}(q) &=& s^{\Delta_a} a_{i}'(q'), \quad \quad \Delta_a = \frac{d+3}{2}, \\
e &=& s^{\Delta_e} e', \quad \quad \Delta_e = \frac{d-3}{2} .
\end{eqnarray}
This scaling analysis indicates that the holon-gauge field interaction vertex becomes marginal at $d = d_{c} = 3$.

Requiring that the spinon-gauge field interaction vertex also has to be marginal at $d = d_{c} = 3$, the scaling dimension of the spinon field is determined as follows
\begin{eqnarray}
\psi_{\sigma \theta}(k) = s^{\Delta_\psi} \psi'_{\sigma \theta}(k') , \quad \quad \Delta_\psi = \frac{d+2}{2} .
\end{eqnarray}
This scaling transformation preserves the linear-dispersion part of spinons while it leads the curvature term to be irrelevant.

Following our recent study \cite{Han_Cho_Kim_SLMQCP}, we consider rescaling of the generalized frequency $\v{K} \rightarrow \sqrt{t \chi^b U} \v{K}$, all dynamical fluctuations of holon $b \rightarrow b/\left \{ (t \chi^b U)^{\frac{d-1}{2}} t \chi^b\right \}^{\frac{1}{2}}$, spinon $\psi \rightarrow \psi / \left \{ (t \chi^b U)^{\frac{d-1}{2}} v_f \right \}^{\frac{1}{2}}$, and U(1) gauge fields $a \rightarrow a/(t \chi^b U)^{\frac{d-1}{4}}$, and both gauge-interaction $e \rightarrow e/ (t \chi^b U)^{\frac{d-1}{4}}$ and holon self-interaction vertices $\lambda \rightarrow (t \chi^b U)^{\frac{d-1}{2}} (t \chi^b)^4 \lambda$.
%
%\begin{eqnarray}
%\v{K} &\rightarrow& \sqrt{t \chi^b U} \v{K} , \nonumber \\
%b &\rightarrow& b/\left \{ (t \chi^b U)^{\frac{d-1}{2}} t \chi^b\right \}^{\frac{1}{2}}, \nonumber \\
%\psi &\rightarrow& \psi / \left \{ (t \chi^b U)^{\frac{d-1}{2}} v_f \right \}^{\frac{1}{2}} , \nonumber \\
%a &\rightarrow& a/(t \chi^b U)^{\frac{d-1}{4}}, \nonumber \\
%e &\rightarrow& e/ (t \chi^b U)^{\frac{d-1}{4}}, \quad \lambda \rightarrow (t \chi^b U)^{\frac{d-1}{2}} (t \chi^b)^4 \lambda,
%\end{eqnarray}
%
As a result, the effective field theory of Eq. (\ref{action1}) becomes more simplified to be
\begin{eqnarray}
S&=& S_f + S_b + S_a + S_{fa} + S_{ba}, \nonumber \\
S_f&=& \int_{k_\theta} \bar{\psi}_{\sigma \theta}
(k_\theta) \left (i \zeta_\psi \boldsymbol{\Gamma} \cdot \v{K} + i \gamma_{d-1} (k_{\theta,d-1} + \kappa k_{\theta,d}^2) \right) \psi_{\sigma \theta}(k_\theta), \nonumber \\
S_b&=& \int_k b_a^\dagger(k) \left (\v{K}^2 + \v{k}^2 \right ) b_a(k), \nonumber \\
&& + \frac{\lambda \mu^\epsilon}{4 N} \int_{k,p,q} b_a^\dagger(k+q) b_a(k) b_{b}^\dagger(p-q) b_b(p) \nonumber \\
S_a&=&\frac{1}{2} \int_{q} a_i(-q)(\zeta_a^2 \v{Q}^2 + \v{q}^2)\delta_{i j} a_j(q), \nonumber \\
S_{fa}&=& \frac{i e \mu^{\frac{\epsilon}{2}}}{\sqrt{N}} \int_{k_\theta,q_\theta} a_\theta(q_\theta) \bar{\psi}_{\sigma \theta}(k_\theta+q_\theta) \gamma_5 \gamma_{d-1} \psi_{\sigma \theta}(k_\theta) \nonumber \\
S_{ba} &=& -\frac{2 e \mu^{\frac{\epsilon}{2}}}{\sqrt{N}} \int_{k,q} k_i a_i(q) b_a^\dagger(k+q) b_a(k) \nonumber \\
&& + \frac{e^2 \mu^{\epsilon}}{N} \int_{k,p,q} a_i (-p+q) a_i (p) b_a^\dagger (k+q) b_a(k) ,
\label{action2}
\end{eqnarray}
where the scale parameter $\mu$ has been introduced to clarify the upper critical dimension of all interaction vertices with $\epsilon = d_{c} - d$. Two anisotropy coefficients for both spinon and gauge-field dynamics are given by $\zeta_\psi = 1/v_f$ and $\zeta_a = (t \chi^b U)^{1/2}$, and the Fermi-surface curvature is $\kappa = t\chi^f / v_f = 1/2k_F$. As pointed out above, the curvature term $\kappa k_d^2$ in the linearized spinon dispersion is irrelevant during the RG process, given by $\kappa < 1/2k_F \ll 1/\Lambda_{a}$, where $\Lambda_{a}$ is the gauge-field momentum cutoff, which will be discussed further later. This originates from the scaling transformation to preserve the holon dispersion in Eq. (\ref{holonscale}). Interestingly, this irrelevant curvature term gives rise to divergence of the Landau damping term much larger than the fermion momentum cutoff $\Lambda_{f}$. The presence of this relevant self-energy in the gauge-field dynamics does not allow us to neglect the Fermi-surface curvature $\kappa$ in the RG transformation, referred to as a dangerously irrelevant operator \cite{Dangerously_Irrelevant_OP}. The key point of the present study beyond the previous investigation is to keep the Fermi-surface curvature $\kappa$ during loop-integrations and to send it to the low-energy limit ($\kappa \rightarrow 0$ with $\kappa < 1/2k_F \ll 1/\Lambda_{a}$) in the final stage of the RG transformation. This extremely overdamped dynamics of U(1) gauge-field fluctuations turns out to cause the fact that all Feynman diagrams including the gauge-field propagator do not show any $1/\epsilon$ poles. This aspect changes the nature of the spin-liquid Mott quantum criticality reported in our previous study \cite{Han_Cho_Kim_SLMQCP} although the emergence of effective one-dimensional spin dynamics remains unchanged.

\subsection{Renormalization group transformation in the one-loop level}

\subsubsection{Renormalized effective action and counter terms}

We perform the perturbative RG analysis based on the effective bare action
\begin{widetext}
\begin{eqnarray}
&&S_B = \int_{k_\theta} \bar{\psi}_{B \sigma \theta}(k_\theta) \left(i \zeta_{B \psi} \boldsymbol{\Gamma} \cdot \v{K}_B + i \gamma_{d-1} \delta_{B k_\theta}\right) \psi_{B \sigma \theta}(k_\theta) + \frac{i e_B}{\sqrt{N}} \int_{k_{B \theta},q_{B \theta}} a_{B \theta}(q_{B \theta}) \bar{\psi}_{B \sigma \theta}(k_{B \theta}+q_{B \theta}) \gamma_5 \gamma_{d-1} \psi_{B \sigma \theta}(k_{B \theta}) \nonumber \\
&&+ \int_{q_B} a_{B i}(-q_B)\left(\zeta_{aB}^2 \v{Q}_B^2 + \v{q}_B^2\right) \delta_{i j} a_{B j}(q_B) \nonumber \\
&& + \int_{k_B} b_{Ba}^\dagger(k_B) \left (\v{K}_B^2 + \v{k}_B^2 \right ) b_{Ba}(k_B) + \frac{\lambda_B }{4 N} \int_{k_B,p_B,q_B} b_{Ba}^\dagger(k_B+q_B) b_{Ba}(k_B) b_{Bb}^\dagger(p_B-q_B) b_{Bb}(p_B) \nonumber \\
&& \nonumber \\ && - \frac{2 e_B}{\sqrt{N}} \int_{k_B,q_B} k_{B i} a_{B i}(q_{B}) b_{Ba}^\dagger(k_B+q_B) b_{Ba}(k)
+ \frac{e_B^2 }{N} \int_{k_B,p_B,q_B} a_{B i} (-p_B+q_B) a_{B i} (p_B) b_{Ba}^\dagger (k_B+q_B) b_{Ba}(k_B) .
\label{bareaction}
\end{eqnarray}
\end{widetext}
Then, various divergent terms appear. These divergent terms can be made to be formally finite, resorting to the dimensional regularization technique. Finally, such divergent terms have to be cancelled by so called counter terms, given by
\begin{widetext}
\begin{eqnarray}
&&S_{CT} = \int_{k_\theta} \bar{\psi}_{\sigma \theta}(k_\theta) \left(A_{\psi 1} i \zeta_{\psi } \boldsymbol{\Gamma} \cdot \v{K} + A_{\psi 2}i \gamma_{d-1} \delta_{k_\theta}\right) \psi_{\sigma \theta}(k_\theta) + A_{\psi a} \frac{i e}{\sqrt{N}} \int_{k_\theta, q_\theta} a_{\theta}(q_\theta) \bar{\psi}_{\sigma \theta}(k_\theta+q_\theta) \gamma_5 \gamma_{d-1} \psi_{ \sigma \theta}(k_\theta) \nonumber \\
&& + \int_{q} a_{i}(-q)\left(A_{a 1} \zeta_{a}^2 \v{Q}^2 + A_{a 2} \v{q}^2\right) \delta_{i j} a_{j}(q) \nonumber \\
&& + \int_{k} b_{a}^\dagger(k) \left ( A_{b 1} \v{K}^2 + A_{b 2} \v{k}^2 \right ) b_{a}(k) + A_{\lambda} \frac{\lambda }{4 N} \int_{k,p,q} b_{a}^\dagger(k+q) b_{a}(k) b_{b}^\dagger(p-q) b_{b}(p) \nonumber \\
&& - A_{ba1} \frac{2 e}{\sqrt{N}} \int_{k,q} k_i a_{i}(q) b_{a}^\dagger(k+q) b_{a}(k)
+ A_{ba2} \frac{e^2 }{N} \int_{k,p,q} a_{i} (-p+q) a_{i} (p) b_{a}^\dagger (k+q) b_{a}(k) . \label{counterterm}
\end{eqnarray}
\end{widetext}
As a result, we find an effective renormalized action $S = S_{B} - S_{CT}$,
\begin{widetext}
%\begin{eqnarray}
%&&S = \int_{k_\theta} \bar{\psi}_{\sigma \theta}(k_\theta) \left(Z_{\psi 1} i \zeta_{\psi } \boldsymbol{\Gamma} \cdot \v{K} + Z_{\psi 2} i \gamma_{d-1} \delta_{k_\theta}\right) \psi_{\sigma \theta}(k_\theta) + Z_{\psi a} \frac{i e}{\sqrt{N}} \int_{k_\theta, q_\theta} a_{\theta}(q_\theta) \bar{\psi}_{\sigma \theta}(k_\theta+q_\theta) \gamma_5 \gamma_{d-1} \psi_{ \sigma \theta}(k_\theta) \nonumber \\
%&& + \int_{q} a_{i}(-q)\left(Z_{a 1} \zeta_{a}^2 \v{Q}^2 + Z_{a 2} \v{q}^2\right) \delta_{i j} a_{j}(q) \nonumber \\
%&& + \int_{k} b_{a}^\dagger(k) \left ( Z_{b 1} \v{K}^2 + Z_{b 2} \v{k}^2 \right ) b_{a}(k) + Z_{\lambda} \frac{\lambda }{4 N} \int_{k,p,q} b_{a}^\dagger(k+q) b_{a}(k) b_{b}^\dagger(p-q) b_{b}(p) \nonumber \\
%&& - Z_{ba1} \frac{2 e}{\sqrt{N}} \int_{k,q} k_i a_{i}(q) b_{a}^\dagger(k+q) b_{a}(k)
%+ Z_{ba2} \frac{e^2 }{N} \int_{k,p,q} a_{i} (-p+q) a_{i} (p) b_{a}^\dagger (k+q) b_{a}(k) , \label{renormalizedaction}
\begin{eqnarray}
&&S = \int_{k_\theta} \bar{\psi}_{\sigma \theta}(k_\theta) \left(i \zeta_{\psi } \boldsymbol{\Gamma} \cdot \v{K} + i \gamma_{d-1} \delta_{k_\theta}\right) \psi_{\sigma \theta}(k_\theta) + \frac{i e}{\sqrt{N}} \int_{k_\theta, q_\theta} a_{\theta}(q_\theta) \bar{\psi}_{\sigma \theta}(k_\theta+q_\theta) \gamma_5 \gamma_{d-1} \psi_{ \sigma \theta}(k_\theta) \nonumber \\
&& + \int_{q} a_{i}(-q)\left(\zeta_{a}^2 \v{Q}^2 + \v{q}^2\right) \delta_{i j} a_{j}(q) \nonumber \\
&& + \int_{k} b_{a}^\dagger(k) \left ( \v{K}^2 + \v{k}^2 \right ) b_{a}(k) + \frac{\lambda }{4 N} \int_{k,p,q} b_{a}^\dagger(k+q) b_{a}(k) b_{b}^\dagger(p-q) b_{b}(p) \nonumber \\
&& - \frac{2 e}{\sqrt{N}} \int_{k,q} k_i a_{i}(q) b_{a}^\dagger(k+q) b_{a}(k)
+ \frac{e^2 }{N} \int_{k,p,q} a_{i} (-p+q) a_{i} (p) b_{a}^\dagger (k+q) b_{a}(k) , \label{renormalizedaction}
\end{eqnarray}
\end{widetext}
which is finite, where all dynamical fields and interaction vertices are renormalized. Here, renormalization coefficients $Z_{r}$ are given by counter-term constants as follows
\begin{eqnarray}
&& Z_{r} = 1 + A_{r} ,
\end{eqnarray}
where $r = \psi 1, ~ \psi 2, ~ \psi a, ~ a 1, ~ a 2, ~ b 1, ~ b 2, ~ \lambda, ~ ba 1, ~ ba 2$. The gauge invariance gives rise to the following Ward identities \cite{Han_Cho_Kim_SLMQCP}
\begin{eqnarray}
&& Z_{\psi 2} = Z_{\psi a} , \quad \quad Z_{b2} = Z_{ba 1} = Z_{ba 2} .
\end{eqnarray}

It is straightforward to see how bare quantities have to be RG-transformed into renormalized ones. First, the momentum relation is given by
\begin{eqnarray}
&&\v{K} = \left (\frac{Z_{b2}}{Z_{b1}}\right )^{\frac{1}{2}} \v{K}_B , \quad \quad \vec{k} = \vec{k}_B .
\end{eqnarray}
Second, the matter-field equation is
\begin{eqnarray}
&& b_{a}(k) = Z_b^{-\frac{1}{2}} b_{B a} (k_B) , \quad \quad Z_b = Z_{b2} \left( \frac{Z_{b2}}{Z_{b1}} \right)^{\frac{d-1}{2}} , \nonumber \\
&& \psi_{\sigma \theta}(k_\theta) = Z_\psi^{-\frac{1}{2}} \psi_{B \sigma \theta}(k_{B \theta}) , \quad \quad Z_\psi = Z_{\psi 2} \left( \frac{Z_{b2}}{Z_{b1}} \right)^{\frac{d-1}{2}} , \nonumber \\ && a_{i}(q) = Z_a^{-\frac{1}{2}} a_{B i}(q_B) , \quad \quad Z_a = Z_{a2}\left( \frac{Z_{b2}}{Z_{b1}} \right)^{\frac{d-1}{2}} . \label{RG_Fields_Integral_Eqs}
\end{eqnarray}
Third, interaction vertices and anisotropy constants are related as
\begin{eqnarray}
&&e_B^2 = e^2 \mu^{\epsilon} Z^{-1}_{a2} \left(\frac{Z_{b2}}{Z_{b1}} \right)^{\frac{d-1}{2}} , \quad \lambda_B = \lambda \mu^{\epsilon} Z_{\lambda} Z_{b2}^{-2} \left( \frac{Z_{b2}}{Z_{b1}}\right)^{\frac{d-1}{2}} , \nonumber \\
&&\zeta_{B \psi}^2 = \zeta_{\psi}^2 \left( \frac{Z_{\psi 1}}{Z_{\psi2}}\right)^2 \frac{Z_{b2}}{Z_{b1}} , \quad \zeta_{Ba}^2 = \zeta_a^2 \frac{Z_{a1}}{Z_{a2}}\frac{Z_{b2}}{Z_{b1}} . \label{RG_Vertex_Integral_Eqs}
\end{eqnarray}

Although we did not show an RG equation involved with the Fermi-surface curvature $\kappa$, where it is irrelevant, we investigate the role of the divergent Landau damping term of the gauge-field dynamics in the RG transformation below. We calculate self-energy and vertex corrections in the one-loop level, where their divergent pieces are identified with counter-term constants $A_{r}$.

\subsubsection{Evaluation of the counter terms in the one-loop level}

\begin{figure*}
\centering
\includegraphics[width=16cm]{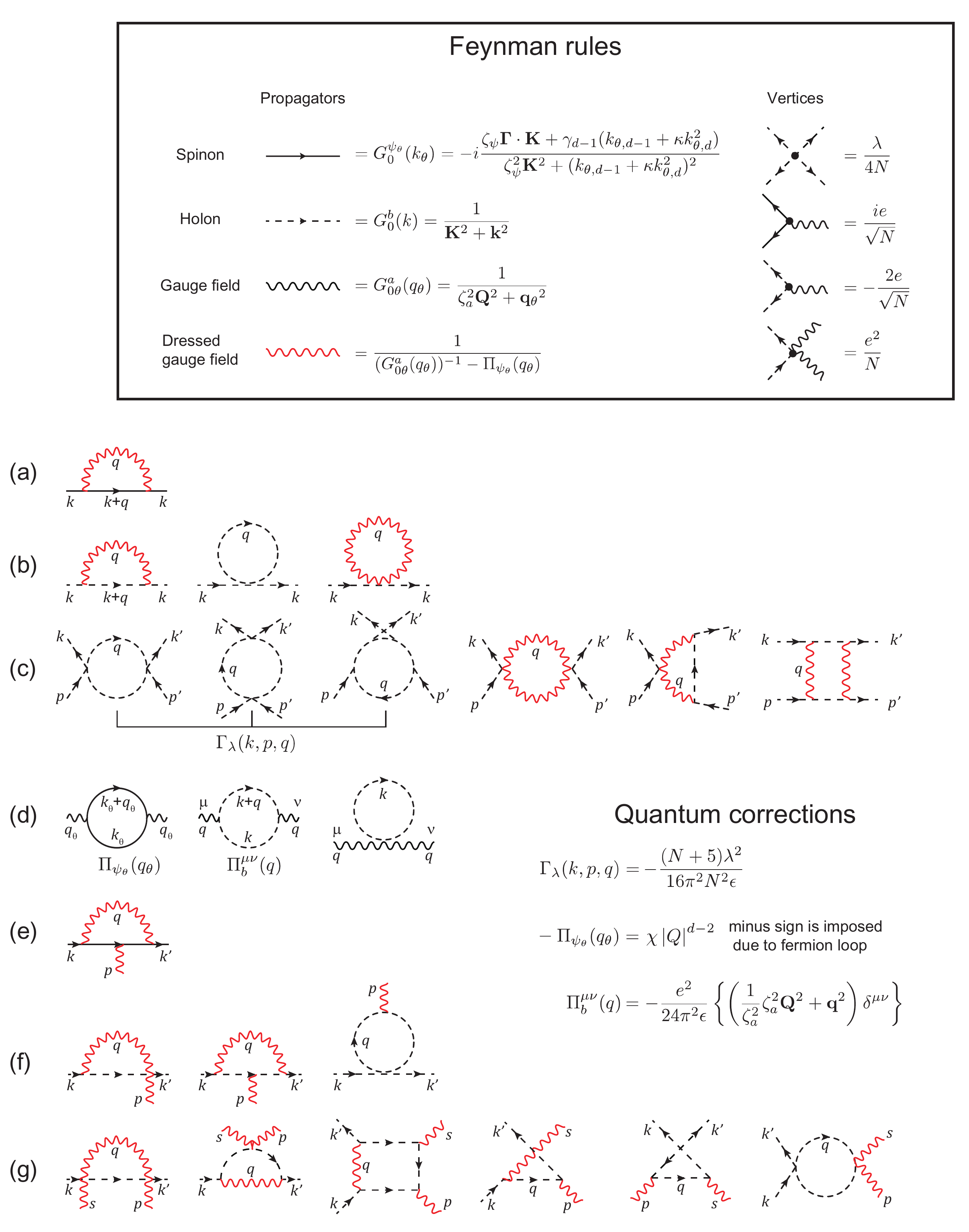}
\caption{Quantum corrections in the one-loop level with Feynman rules. (a) Spinon self-energy corrections in $S_{f}$ of Eq. (\ref{action2}). (b) Holon self-energy corrections in $S_{b}$ of Eq. (\ref{action2}). (c) Effective self-interaction $\lambda$-vertex corrections in $S_{b}$ of Eq. (\ref{action2}). (d) Gauge-field self-energy corrections in $S_{a}$ of Eq. (\ref{action2}). (e) Spinon-gauge field interaction vertex corrections in $S_{fa}$ of Eq. (\ref{action2}). (f) Holon-gauge field interaction vertex corrections for $\langle a b^\dagger b \rangle$ in $S_{ba}$ of Eq. (\ref{action2}). (g) Holon-gauge field interaction vertex corrections for $\langle a a b^\dagger b \rangle$ in $S_{ba}$ of Eq. (\ref{action2}). Red wavy lines represent the renormalized gauge-field propagator Eq. (\ref{Overdamping_Gauge_Field}), where the extremely overdamped dynamics of U(1) gauge-field fluctuations does not cause any renormalization effects, thus neglected.}
\label{oneloops}
\end{figure*}

We introduce propagators of spinons, holons, and gauge-field fluctuations,
\begin{eqnarray}
G_0^{\psi_\theta}(k_\theta) &=& -i \frac{\zeta_\psi \boldsymbol{\Gamma} \cdot \v{K} + \gamma_{d-1} (k_{\theta,d-1} + \kappa k_{\theta,d}^2)}{\zeta_\psi^2 \v{K}^2 + (k_{\theta,d-1} + \kappa k_{\theta,d}^2)^2},\\
G_0^b(k) &=& \frac{1}{\v{K}^2 + \v{k}^2}, \\
G_{0 i j}^a(q) &=& \frac{\delta_{i j}}{\zeta_a^2 \v{Q}^2 + \v{q}^2} ~ \text{ for the holon-gauge field vertex} , \nonumber  \\
G_{0 \theta}^a(q_\theta) &=& \frac{1}{\zeta_a^2 \v{Q}^2 + \v{q_\theta}^2} ~ \text{ for the spinon-gauge field vertex} , \nonumber  \\ \label{apropagator}
\end{eqnarray}
where Feynman rules are shown in Fig. \ref{oneloops}. Based on such Feynman rules, we perform the perturbative analysis for all interaction vertices, where all possible quantum corrections in the one-loop level are shown as Feynman diagrams of Fig. \ref{oneloops}.

First, we evaluate the gauge-field self-energy correction, given by the spinon-polarization bubble diagram as shown in the first diagram of Fig. \ref{oneloops} (d),
%
%\begin{eqnarray}
%- \Pi_{\psi_\theta} (q_\theta) &=& 2 \left(\frac{i e \mu^{\frac{\epsilon}{2}}}{\sqrt{N}}\right)^2 \int \frac{d^{d+1}k_\theta}{(2 \pi)^{d+1}} \nonumber \\
%&&\times \text{tr}[G_0^{\psi_\theta}(k_\theta) \gamma_5 \gamma_{d-1} G_0^{\psi_\theta} (k_\theta+q_\theta) \gamma_5 \gamma_{d-1}] \nonumber \\
%&=& \chi \frac{e^2 \mu^\epsilon}{N} \frac{\abs{Q}^{d-1}}{\abs{q_{\theta,d}}} ~ \text{ in the $\theta-$rotated coordinate} \nonumber \\
%&=& \chi \frac{e^2 \mu^\epsilon}{N} \frac{\abs{Q}^{d-1}}{\abs{\v{q}}} ~ \text{ in the global coordinate} ,
%\end{eqnarray}
%
\begin{widetext}
\begin{eqnarray}
- \Pi_{\psi_\theta} (q_\theta) &=& 2 \left(\frac{i e \mu^{\frac{\epsilon}{2}}}{\sqrt{N}}\right)^2 \int \frac{d^{d+1}k_\theta}{(2 \pi)^{d+1}}  \text{tr}[G_0^{\psi_\theta}(k_\theta) \gamma_5 \gamma_{d-1} G_0^{\psi_\theta} (k_\theta+q_\theta) \gamma_5 \gamma_{d-1}] \nonumber \\
&=& 2 \left (\frac{i e \mu^{\frac{\epsilon}{2}}}{\sqrt{N}} \right)^2 \int \frac{dk_{\theta,d-1}dk_{\theta,d} d\v{K}}{(2 \pi)^{d+1}}\frac{\v{K} \cdot (\v{K}+\v{Q})-\delta_{k_\theta}\delta_{k_\theta+q_\theta}}{(\v{K}^2+\delta_{k_\theta}^2)((\v{K}+\v{Q})^2+\delta_{k_\theta+q_\theta}^2)} \nonumber \\
&=&2 \left (\frac{i e \mu^{\frac{\epsilon}{2}}}{\sqrt{N}} \right)^2 \int \frac{d\v{K}}{(2 \pi)^{d+1}}\left( \frac{\v{K} \cdot (\v{K}+\v{Q})}{\abs{\v{K}} \abs{\v{K}+\v{Q}}}-1 \right) \frac{\arctan(\frac{2 \kappa q_d k_d}{\abs{\v{K}+\v{Q}}+\abs{\v{K}}})}{4 \kappa q_d} \Bigg ]^{k_d=\Lambda_d}_{k_d = -\Lambda_d} \nonumber \\
&\sim& 2 \left (\frac{i e \mu^{\frac{\epsilon}{2}}}{\sqrt{N}} \right)^2 \int \frac{d\v{K}}{(2 \pi)^{d+1}} \left( \frac{\v{K} \cdot (\v{K}+\v{Q})}{\abs{\v{K}} \abs{\v{K}+\v{Q}}}-1 \right) \frac{\Lambda_d}{\abs{\v{K}+\v{Q}}+\abs{\v{K}}} \nonumber \\
&=& \chi \abs{Q}^{d-2}. \label{chi}
\end{eqnarray}
\end{widetext}
Here, the minus sign in $\Pi_{\psi_\theta} (q_\theta)$ results from the fermion loop and the divergent coefficient $\chi$ of the final result is proportional to the UV cutoff $\Lambda_d$ along the Fermi surface. We perform the integration of $k_{\theta, d-1}$ up to infinity in the second line while the momentum integration of $k_d$ is taken into account up to the UV cutoff $\Lambda_d$, not infinity. This integration domain originates from the Fermi-surface constraint with an irrelevant curvature coefficient $\kappa$, given by $\kappa < 1/2k_F \ll 1/\Lambda_{a}=1/\Lambda_{d}$. See Fig. \ref{patchconstruction} (a) and (b). This Fermi-surface constraint allows us to take an approximation $\frac{\arctan(k x)}{x} \sim k$ for $x \ll 1$ in the third line. The self-energy correction has been evaluated in the local patch coordinate (the second form of Eq. (\ref{apropagator})), but this expression can be trivially translated into that of the global coordinate (the first form of Eq. (\ref{apropagator})), where this expression depends on only frequency. This correction term for the gauge-field dynamics has a different form from $\Pi_{\psi_\theta} \propto \frac{1}{\kappa} \frac{\abs{Q}^{d-1}}{\abs{\v{q}}}$ of the spin-liquid state \cite{Dimensional_Regularizarion_FS_I,Dimensional_Regularizarion_FS_II}, which results from the RG flow of the Fermi-surface curvature $\kappa$. $\kappa$ is marginal in the spin-liquid phase while it is irrelevant at this spin-liquid Mott quantum criticality. As a result, this Landau damping term diverges at the spin-liquid Mott transition.

Now, we introduce this nonlocal term into the gauge-field propagator as
\begin{eqnarray}
&& D^a(q_\theta) = \frac{1}{(G_{0 \theta}^a (q_\theta))^{-1} - \Pi_{\psi_\theta}(q_\theta)} , \label{Overdamping_Gauge_Field}
\end{eqnarray}
given by red wavy lines in all Feynman diagrams of Fig. \ref{oneloops}. Then, we calculate the spinon's self-energy correction, shown in the first diagram of Fig. \ref{oneloops} (a),
\begin{eqnarray}
\Sigma_{\psi_\theta} (k_{\theta}) &=& \left( \frac{i e \mu^{\frac{\epsilon}{2}}}{\sqrt{N}} \right)^2 \int_{q_{\theta}} \gamma_5 \gamma_{d-1} G_0^{\psi_{\theta}} (k_{\theta}+q_{\theta}) \gamma_5 \gamma_{d-1} D^a(q_{\theta}) \nonumber \\
&=& \frac{e^2}{N(2 \pi)^{d+1}}[A i \zeta_\psi \boldsymbol{\Gamma} \cdot \v{K} + B i \gamma_{d-1} k_{\theta, d-1}] \nonumber \\
&& \times (\ln{(\Lambda_{d-1}^2 + \chi^2)} - \ln{\chi^2}) + O(\epsilon^0) . \label{spinonself}
\end{eqnarray}
Here, $A$ and $B$ are functions of $\zeta_\psi$ and $\zeta_a$, which are finite and not shown. Although we keep the coefficient $\chi$ to be finite during the loop integration, we point out that it is proportional to $\Lambda_d$ as shown in Eq. (\ref{chi}). As a result, the log divergence of $\ln{(\Lambda_{d-1}^2 + \chi^2)}$ has to be cancelled by that of $\ln{\chi^2}$ in Eq. (\ref{spinonself}) in the following way
\begin{eqnarray}
&& \ln{(\Lambda_{d-1}^2 + \chi^2)} - \ln{\chi^2}
\sim \ln{(\Lambda_{d-1}^2/\Lambda_d^2+1)} \ll 1 ,
\end{eqnarray}
where $\Lambda_d \gg \Lambda_{d-1}$ has been utilized. We recall Fig. \ref{patchconstruction} (a) and (b), where $\Lambda_{f}$ and $\Lambda_a$ corresponds to $\Lambda_{d-1}$ and $\Lambda_d$, respectively. Similarly, all Feynman diagrams including the dressed gauge-field propagator do not show $1/\epsilon$ pole corrections.

Finally, we evaluate other quantum corrections without U(1) gauge-field fluctuations. Those are the second diagram in Fig. \ref{oneloops} (b) denoted as $\Sigma_{b,2}$, the first three diagrams in Fig. \ref{oneloops} (c) contributing to $\Gamma_{\lambda}$, the second and third diagrams in Fig. \ref{oneloops} (d) represented as $\Pi^{i j}_{b}$ and $\Sigma_{a}$, respectively, the third diagram in Fig. \ref{oneloops} (f) given by $\Gamma^i_{ab^\dagger b,3}$, and the last two diagrams in Fig. \ref{oneloops} (g) expressed as $\Gamma_{aab^\dagger b,5}$ and $\Gamma_{aab^\dagger b,6}$, respectively.

First, $\Sigma_{b,2}$ and $\Sigma_a$ are both proportional to $\int_k G^b_0 (k)$. This does not cause any $1/\epsilon$ pole contribution at quantum criticality while it can give rise to the mass renormalization in the holon dynamics.

Second, $\Gamma^i_{ab^\dagger b,3}$ has no $1/\epsilon$ pole correction, where its loop integral is UV finite at $d=3-\epsilon$ as follows
\begin{eqnarray}
\Gamma^i_{ab^\dagger b,3} &=& \int_k N \frac{\lambda \mu^{\epsilon}}{4 N} \frac{2 e \mu^{\epsilon}}{\sqrt{N}} k^i G_0^b(k) G_0^b(k+q) \nonumber \\
&=& O(\epsilon^0).
\end{eqnarray}

Third, $\Gamma_{aab^\dagger b,5}$ and $\Gamma_{aab^\dagger b,6}$ are both UV divergent, but their sum is finite, given by
\begin{eqnarray}
&& \Gamma_{aab^\dagger b,5} + \Gamma_{aab^\dagger b,6} \nonumber \\
&=& N \frac{\lambda \mu^{\epsilon}}{4 N} \left (-\frac{2 e^{\frac{\epsilon}{2}}}{\sqrt{N}} \right )^2 \int_k k_i k_i G^b_0(k) G^b_0(k+q) G^b_0(k+p) \nonumber \\
&& + N \frac{\lambda \mu^{\epsilon}}{4 N} \frac{e^2 \mu^{\epsilon}}{N} \int_k G^b_0(k) G^b_0(k+q) \nonumber \\
&=& O(\epsilon^0).
\end{eqnarray}

Now, the remaining diagrams are $\Gamma_\lambda$ and $\Pi^{i j}_b$. Holon self-interaction vertex corrections $\Gamma_\lambda$ are given by
\begin{eqnarray}
\Gamma_\lambda(k,p,q) &=& - \frac{N+5}{2}\left( \frac{\lambda \mu^\epsilon}{N}\right)^2 \int \frac{d^{d+1} k'}{(2 \pi)^{d+1}} G_0^b(k') G_0^b(k'+q) \nonumber \\
&=& - \frac{(N+5)\lambda^2}{16 \pi^2 N^2 \epsilon} + O(\epsilon^0) ,
\end{eqnarray}
typical for self-interacting bosons \cite{RG_Typical_Phi4}.

Finally, the holon polarization function $\Pi^{i j}_b$ is given by
\begin{eqnarray}
\Pi_b^{i j}(q) &=& N\left (-\frac{2 e \mu^{\frac{\epsilon}{2}}}{\sqrt{N}} \right )^2 \int \frac{d^{d+1} k }{(2 \pi)^{d+1}} k^i k^j G_0^b(k+q)  G_0^b(k) \nonumber \\
&=& -\frac{e^2}{24 \pi^2 \epsilon} \left \{ \left (\frac{1}{\zeta_a^2} \zeta_a^2 \v{Q}^2 + \v{q}^2 \right ) \delta^{\mu \nu} \right \} + O(\epsilon^0) ,
\end{eqnarray}
which renormalizes the gauge-field dynamics at this Mott quantum criticality \cite{RG_Typical_Phi4}.

As a result, we find all the counter terms as follows
\begin{eqnarray}
A_{b1} &=& A_{b2} = A_{\psi 1} = A_{\psi 2} = A_{\psi a} = A_{ba1} = A_{ba2} = 0, \nonumber \\
A_{a1}&=& -\frac{e^2}{24 \pi^2 \zeta_a^2 \epsilon}, \quad A_{a2} = -\frac{e^2}{24 \pi^2 \epsilon}, \nonumber \\
A_\lambda &=& \frac{(N+5)\lambda}{16 \pi^2 N \epsilon} , \label{CT_Results}
\end{eqnarray}
where renormalization effects are given by critical holon excitations only.

\subsubsection{Renormalization group $\beta-$functions for coupling constants and Callan-Symanzik equations for correlation functions}

\begin{figure}
\centering
\includegraphics[width=8cm]{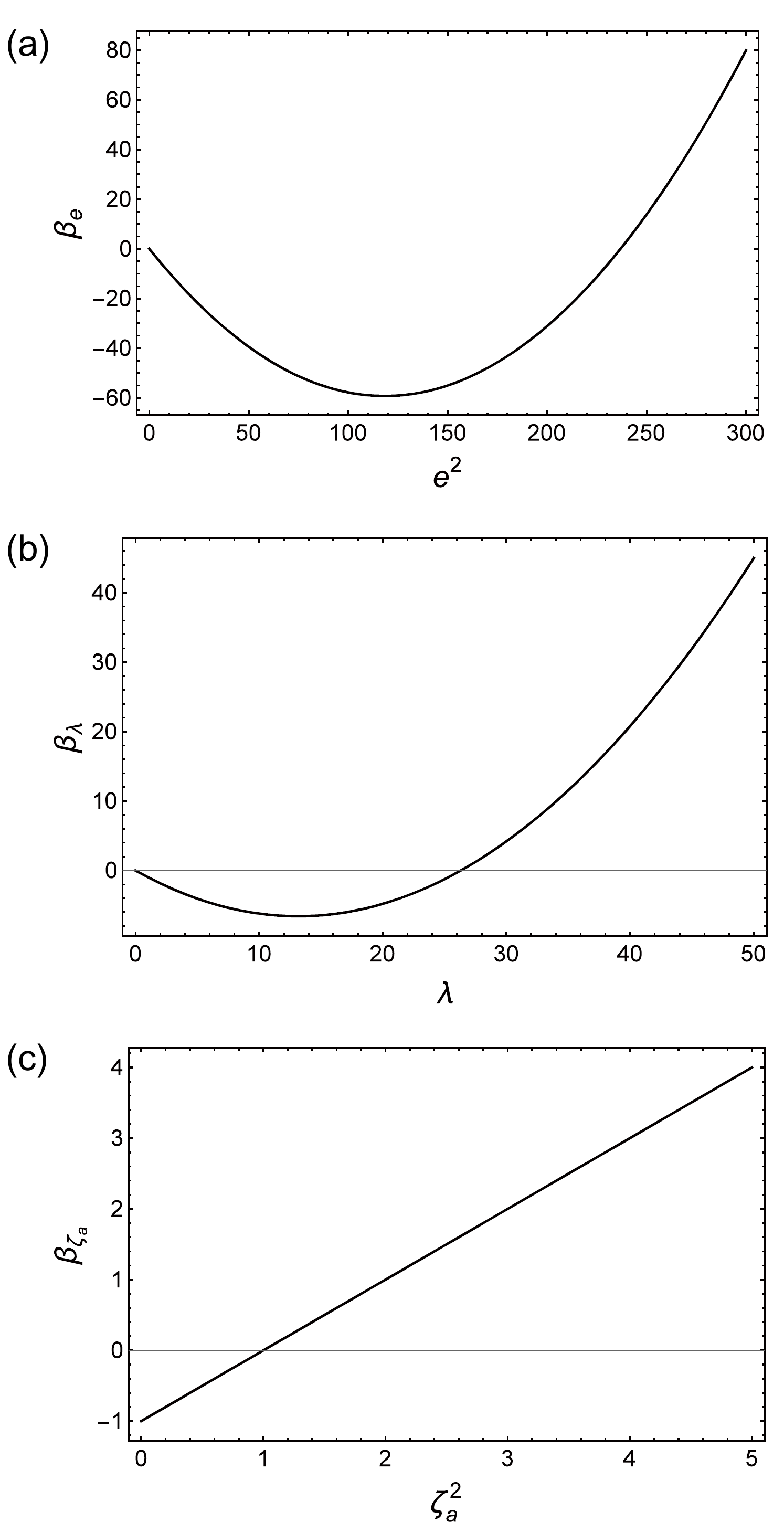}
\caption{$\beta-$functions of (a) the gauge-field coupling constant $e$, (b) the self-interaction coupling constant $\lambda$, and (c) the anisotropy constant $\zeta_a$ in the gauge-field dispersion. Although the gauge-field coupling constant remains to be finite at the spin-liquid Mott quantum critical point, the quantum critical charge dynamics belongs to the XY universality class, described by the Wilson-Fisher fixed point \cite{RG_Textbook,RG_Typical_Phi4}.}
\label{betafunctions}
\end{figure}

RG equations (\ref{RG_Vertex_Integral_Eqs}) can be translated into the following differential equations
\begin{eqnarray}
\frac{\mu}{e^2} \frac{de^2}{d\mu} &=& -\epsilon + \frac{\mu}{Z_{a2}} \frac{dZ_{a2}}{d\mu} , \\
\frac{\mu}{\lambda} \frac{d \lambda}{d\mu} &=& -\epsilon - \frac{\mu}{Z_{\lambda}} \frac{dZ_{\lambda}}{d\mu} , \\
\frac{\mu}{\zeta_a^2} \frac{d \zeta_a^2}{d\mu} &=& \frac{\mu}{Z_{a2}/Z_{a1}} \frac{dZ_{a2}/Z_{a1}}{d\mu} ,
\end{eqnarray}
where $\zeta_\psi^{2}$ does not renormalize due to the extremely overdamped dynamics of U(1) gauge fields. Introducing the counter-term coefficients of Eq. (\ref{CT_Results}) into the above with $\epsilon=1$, we find the RG $\beta-$functions of the coupling constants as follows
\begin{eqnarray}
\beta_e \equiv \mu \frac{d e^2}{d \mu} &=& e^2 \left (-1+ \frac{e^2}{24 \pi^2} \right) , \\
\beta_\lambda \equiv \mu \frac{d \lambda}{d \mu} &=& \lambda \left (-1+ \frac{(N+5)\lambda}{16 \pi^2 N} \right ) , \\
\beta_{\zeta_a} \equiv \mu \frac{d \zeta_a^2}{d \mu} &=& \frac{e^2}{24 \pi^2} \left ( \zeta_a^2 - 1 \right ) ,
\end{eqnarray}
where $\beta_{\zeta_\psi} = 0$. The spin-liquid Mott quantum critical point is identified with a fixed point at $e^2=e_*^2 = 24 \pi^2$, $\lambda^2= \lambda_*^2 = 16\pi^2/6 $, and $\zeta_a = \zeta_{a*} = 1$ as shown in Fig. \ref{betafunctions}.

We emphasize that this fixed point differs from that of our previous study \cite{Han_Cho_Kim_SLMQCP}, where renormalization effects also occur from gauge-field fluctuations. In this study, we have shown that the role of dangerously irrelevant operator kills such renormalization effects from gauge-field fluctuations, resulting in unexpectedly simple mean-field type dynamics at the Mott quantum criticality. Actually, anomalous scaling dimensions of both matter and gauge fields are different from those of the previous study. As a result, the quantum critical charge dynamics belongs to the XY universality class \cite{RG_Textbook,RG_Typical_Phi4} although U(1) gauge-field fluctuations acquire anomalous scaling dynamics from critical charge fluctuations in the one-loop level.

To calculate correlation functions near the Mott quantum critical point, we introduce the Callan-Symanzik equation \cite{RG_Textbook,RG_Typical_Phi4}, given by \cite{Han_Cho_Kim_SLMQCP}
\begin{eqnarray}
 && [z \v{K}_i \cdot \grad_{\v{K}_i} + \vec{k}_i \cdot \grad_{\vec{k}_i} - \beta_{e} \pd{}{e^2} - \beta_{\lambda} \pd{}{\lambda} - \beta_{\zeta_a} \pd{}{\zeta_a} \nonumber \\ && - 2m (-\frac{5-\epsilon}{2} + \eta_{\psi}) - 2n (-\frac{6-\epsilon}{2} + \eta_{b}) - 2l (-\frac{6-\epsilon}{2} + \eta_{a}) \nonumber \\
 && - \{ z (2-\epsilon) + 2 \} ] ~ G^{(m,n,l)} ({k_i}; e,\lambda,\zeta_a,\mu) = 0 ,
\end{eqnarray}
where the renormalized $(m + n + l)-$point Green's function is
\begin{eqnarray}
 && G^{(m,n,l)} ({k_i}; e,\lambda,\zeta_a,\mu) ~ \delta^{(d+1)}(\{ k_{i} \}) \nn && = \Big\langle \bar{\psi}(k_{1}) \cdot \cdot \cdot \bar{\psi}(k_{m}) \psi(k_{m+1}) \cdot \cdot \cdot \psi(k_{2m}) \nn && b^{\dagger}(k_{2m+1}) \cdot \cdot \cdot b^{\dagger}(k_{2m+n}) b(k_{2m+n+1}) \cdot \cdot \cdot b(k_{2m+2n}) \nn && a(k_{2m+2n+1}) \cdot \cdot \cdot a(k_{2m+2n+2l}) \Big\rangle .
\end{eqnarray}
Here, the coordinate transformation has been assumed appropriately. The dynamical critical exponent $z$ and the anomalous scaling dimension of each dynamical field are
\begin{eqnarray}
z &=& 1 - \frac{1}{2} \frac{\mu}{Z_{b2}/Z_{b1}} \frac{\partial Z_{b2}/Z_{b1}}{\partial \mu} , \nn \eta_{\psi} &=& \frac{1}{2} \frac{\mu}{Z_{\psi}} \frac{\partial Z_{\psi}}{\partial \mu} , ~~~ \eta_{b} = \frac{1}{2} \frac{\mu}{Z_{b}} \frac{\partial Z_{b}}{\partial \mu} , ~~~ \eta_{a} = \frac{1}{2} \frac{\mu}{Z_{a}} \frac{\partial Z_{a}}{\partial \mu} , \nn
\end{eqnarray}
where the field renormalization constants of $Z_{\psi}$, $Z_{b}$, and $Z_{a}$ are defined in Eq. (\ref{RG_Fields_Integral_Eqs}). One can derive this Callan-Symanzik equation, constructing an RG equation between the bare and renormalized Green's functions based on the RG equations for fields and interaction vertices. We refer this derivation to the previous study \cite{Han_Cho_Kim_SLMQCP}.

Focusing on the spin-liquid Mott critical fixed point, given by $z_*=1$, $\beta_{e_*} = \beta_{\lambda_*} = 0$, $\eta_{\psi*} = \eta_{b*} = 0$, and $\eta_{a*} = 1/2 $, we obtain
%
%\begin{eqnarray}
%\eta_a &=& \frac{1}{2} \frac{\mu}{Z_a} \pd{Z_a}{\mu}, \\
%Z_a &=& 1+A_a = 1 - \frac{e^2}{24 \pi^2 \epsilon}.
%\end{eqnarray}
%
\begin{eqnarray}
 && [\v{K}_i \cdot \grad_{\v{K}_i} + \vec{k}_i \cdot \grad_{\vec{k}_i} + 4m + 5n + 5l - 3 \nn && - 2l \eta_{a*}] ~ G^{(m,n,l)} (k_i;e_*,\lambda_*,\zeta_{a*},\mu) = 0 .
\end{eqnarray}
Solving this equation, we find one-particle Green's functions as follows
\begin{eqnarray}
&& \Big\langle \bar{\psi}_{\sigma \theta}(\v{K},k_{\theta,d-1}) \psi_{\sigma \theta}(\v{K},k_{\theta,d-1}) \Big\rangle \nn &=& \frac{1}{|k_{\theta,d-1}|^{2 - z_* - 2 \eta_{\psi*}}} F_{\psi_{\sigma \theta}}\Big(\frac{|\v{K}|^{1/z_*}}{|k_{\theta,d-1}|}\Big) \propto \frac{1}{|k_{\theta,d-1}|}
\end{eqnarray}
for spinons,
\begin{eqnarray}
&& \Big\langle b^\dagger (\v{K},\v{k}) b (\v{K},\v{k}) \Big\rangle = \frac{1}{|\v{k}|^{2 - 2 \eta_{b*}}} F_{b}\Big(\frac{|\v{K}|}{|\v{k}|}\Big) \propto  \frac{1}{|\v{k}|^2}
\end{eqnarray}
for holons,
\begin{eqnarray}
&& \Big\langle a(-\v{K},-\v{k}) a(\v{K},\v{k}) \Big\rangle = \frac{1}{|\v{k}|^{2 - 2 \eta_{a*}}} F_{a}\Big(\frac{|\v{K}|}{|\v{k}|}\Big) \propto  \frac{1}{|\v{k}|}
\end{eqnarray}
for U(1) gauge fields. Here, $F_{\psi_{\sigma \theta}}\Big(\frac{|\v{K}|^{1/z_*}}{|k_{\theta,d-1}|}\Big)$ and $F_{b, a} \Big(\frac{|\v{K}|}{|\v{k}|}\Big)$ are nonsingular scaling functions, which can be found by direct computations.

It is interesting to observe that both matter fields of spinons and holons do not acquire any anomalous scaling dimensions at the Wilson-Fisher-type fixed point, where the gauge-field coupling constant remains to be finite and different from the Wilson-Fisher fixed point \cite{RG_Textbook,RG_Typical_Phi4}. $\eta_{b*} = 0$ has to be considered as an artifact of the one-loop level RG analysis, where effective self-interactions between critical charge fluctuations give rise to an anomalous scaling dimension in the two-loop level \cite{RG_Textbook,RG_Typical_Phi4} at least. On the other hand, $\eta_{\psi*} = 0$ is expected to hold beyond the one-loop RG analysis since quantum corrections from U(1) gauge-field fluctuations do not cause any renormalization effects.

\begin{table}
\begin{tabular}{c @{$\quad$ $\quad$ $\quad$ } c @{$\quad$ $\quad$ $\quad$ } c c c}
\\
\hline \hline & $N=1$ & $N$ $\rightarrow$ $\infty$\\
\hline \\
$e_*^2/4 \pi^2$ & 6 & 6 \\
$\lambda_*^2/4 \pi^2$ & 2/3 & 4 \\
$\zeta_{a*}^2 $ & 1 & 1 \\
$z_*$ & 1 & 1 \\
$\eta_{\psi *}$ & 0 & 0 \\
$\eta_{b *}$ & 0 & 0 \\
$\eta_{a *}$ & 0.5 & 0.5 \\
\hline \hline \\
\end{tabular}
\caption{Fixed point values, critical exponents, and anomalous scaling dimensions.}
\end{table}

\section{Conclusion} \label{Conclusion}

\subsection{Summary and discussion}

In the present study, we revisited the spin-liquid Mott quantum criticality for $\kappa-(BEDT-TTF)_{2}Cu_{2}(CN)_{3}$ \cite{Kanoda_MQCP_Transport}. The previous study \cite{Han_Cho_Kim_SLMQCP} has claimed that the spin dynamics shows effectively one-dimensional Luttinger-liquid physics \cite{Luttinger_Liquid_Textbook} while the quantum critical charge dynamics belongs to the $(2+1)-$dimensional inverted XY (IXY) universality class \cite{IXY_Review}. However, our present study revealed that the role of U(1) gauge-field fluctuations in the spin-liquid Mott quantum criticality of the previous study had been overestimated because their extremely overdamped dynamics had not been taken into account in the RG analysis. It turns out that the generic scale invariance \cite{Generic_Scale_Invariance} gives rise to a dangerously irrelevant operator \cite{Dangerously_Irrelevant_OP}, here, the spinons' Fermi-surface curvature term, which results in divergence of the self-energy term for U(1) gauge-field fluctuations. As a result, such extremely overdamped gauge-field dynamics does not cause any renormalization effects to both spin and critical charge dynamics. This leads us to conclude that the critical spin dynamics still shows one-dimensional Luttinger-liquid physics while the quantum critical charge dynamics belongs to the XY universality class \cite{RG_Textbook,RG_Typical_Phi4} instead of the IXY. Here, we have to point out that the Luttinger-liquid-type spin dynamics may result from residual effective interactions between spinons, irrelevant in the presence of the spinon Fermi surface at UV but marginal in the effectively one-dimensional dynamics at IR.

We would like to emphasize that the generic scale invariance is ubiquitous in quantum criticality of metals, well discussed in Ref. \cite{Generic_Scale_Invariance}, where the appearance of dangerously irrelevant operators in the presence of generic scale invariance has been pointed out. In particular, symmetry breaking quantum criticality in disordered metals has been discussed, which results in an exotic mean-field type behavior modified by a dangerously irrelevant operator. In this respect the conclusion of the present study is in parallel with this general perspective at quantum criticality.

To verify the one-dimensional spin dynamics, it is necessary to investigate not only the uniform spin susceptibility but also the $2 k_{F}$ spin susceptibility, where $k_{F}$ is the Fermi momentum of spinon excitations. It is natural to expect that both spin susceptibilities will show divergences in the DMFT framework since critical local magnetic-moment fluctuations affect both momentum channels. On the other hand, only the $2 k_{F}-$channel spin susceptibility diverges in the Luttinger-liquid spin dynamics, where the uniform spin susceptibility vanishes in a power-law fashion as a function of temperature, which originates from vanishing density of states \cite{Luttinger_Liquid_Textbook}. We recall that only the uniform spin susceptibility diverges as a power-law fashion of temperature in the U(1) spin-liquid state \cite{Nagaosa_Lee_SL} while the $2 k_{F}$ spin susceptibility dies out, both of which originate from U(1) gauge-field fluctuations \cite{Dimensional_Regularizarion_FS_I}.

We speculate that the bifurcation behavior of the electrical resistivity near the Mott transition \cite{Kanoda_MQCP_Transport} can be explained within the XY universality class of critical charge dynamics. Frankly speaking, it is not easy to calculate the electrical resistivity based on the effective field theory Eq. (\ref{action2}). In particular, we suspect that the Ioffe-Larkin composition rule for response functions \cite{Nagaosa_Lee_SL} may not work due to extremely overdamped dynamics of U(1) gauge fields. Suppose that the electrical resistivity is given by the holon transport coefficient only, which results from the gauge-field dynamics. Then, the metal-insulator transition is nothing but the superfluid-insulator transition in the holon sector. We claim that the superfluid to Mott insulator transition gives rise to the bifurcation behavior in the electrical resistivity near the quantum critical point \cite{SF_Mott_Transition_Review_I,SF_Mott_Transition_Review_II}. In this case the critical exponent is given by that of the XY universality class, expected to be $\nu z \approx 0.67$, where $\nu \approx 0.67$ is the correlation-length exponent and $z = 1$ is the dynamical critical exponent.

\subsection{Future perspectives}

Recently, a fracton spin-liquid phase has been discussed both extensively and intensively \cite{Fracton_Review_I,Fracton_Review_II}. In this respect it is natural to consider a quantum phase transition from a fracton-type spin-liquid Mott insulating state to a Landau's Fermi-liquid phase. Here, either U(1) tensor gauge theory or more stable Z$_{2}$ tensor gauge theory possibly appears to describe the corresponding fracton spin-liquid state. To discuss this exotic Mott transition, one may introduce charge-fluctuation dynamics into the prototype lattice model for the fracton phase, for example, X$-$cube model \cite{Xcube_fracton} or Haah's cubic code model \cite{Haah_fracton}. Resorting to the parton construction \cite{Fracton_Review_I,Fracton_Review_II} which takes into account not only the spin dynamics but also the charge dynamics, one can reformulate the effective lattice model Hamiltonian in terms of fractionalized excitations, where conventional gauge fluctuations and spinon excitations in the present study are replaced with tensor-type gauge-field fluctuations and fracton-like spinon excitations. Then, one may perform the RG analysis for this fracton Mott quantum criticality, regarded to be an interesting research direction.

However, it is much more exciting to speculate that the exotic fracton physics may appear in the present spin-liquid Mott quantum criticality. This remarkable but seemingly unrealistic perspective starts from the following observation that the so called dimensional reduction occurs in the spinon dynamics from $2d$ to $1d$ in the vicinity of the Mott transition. One may regard this emergent one-dimensional spin dynamics as a certain limit of nematicity, where the spinon Fermi surface elongates to break C$_{4}$ symmetry and results in localization along the elongation direction. This would be analogous to the fact that the Rashba-type spin-orbit coupled model describes the surface state of a topological insulating phase when the strength of the Rashba spin-orbit coupling constant becomes infinite. Such quadrupolar fluctuations occur quite commonly, maybe ubiquitously, in the flat-band system, for example, fractional quantum Hall liquids, magic angle twisted bilayer graphene, and etc. \cite{Nematicity_Review} Here, an idea is as follows: When these quadrupolar fluctuations are coupled to critical charge fluctuations, a conservation law may newly appear at low energies, that is, the emergent conservation law of dipoles in the spin dynamics. In this respect the present RG analysis gives rise to interesting insight for the possible connection to the emergent fracton-like behavior in the spin dynamics near the Mott quantum criticality. It would be an exceptional research direction to find a fixed-point effective field theory of this fracton Mott quantum criticality in the perspective of the emergent fracton dynamics coupled to critical charge fluctuations. The coupled wire construction \cite{Coupled_Wire_Construction_Fracton} may shed light on this direction, i.e., showing the emergence of the dipolar conservation law at a novel fixed point in the vicinity of the fracton spin-liquid Mott quantum criticality.

The original motivation of the present study is to show how the U(1) spin-liquid phase with a spinon Fermi surface becomes destabilized by critical charge fluctuations. Unfortunately, we cannot find a way to perform the RG analysis near this spin-liquid fixed point, which also results from the generic scale invariance. This point has been well discussed in our previous study \cite{Han_Cho_Kim_SLMQCP}. Here, we investigated one possible spin-liquid Mott quantum criticality near the Wilson-Fisher fixed point of critical charge dynamics. In this case it is not easy to see how the dimensional reduction occurs in the spin dynamics, and thus, it is not possible to figure out how the fracton dynamics appears possibly at the Mott transition. It is necessary to develop how to perform the RG analysis near the spin-liquid fixed point.

\section*{Acknowledgement}

This work was supported by the Ministry of Education, Science, and Technology (No. 2011-0030046) of the National Research Foundation of Korea (NRF).


\begin{thebibliography}{9}
\bibitem{Organic_Mott_Ins_Review} B. J. Powell and R. H. McKenzie, \textit{Quantum frustration in organic Mott insulators: from spin liquids to unconventional superconductors}, Rep. Prog. Phys. {\bf 74}, 056501 (2011).
\bibitem{Kappa_BEDT_NMR}  Y. Shimizu, K. Miyagawa, K. Kanoda, M. Maesato and G. Saito, \textit{Spin Liquid State in an Organic Mott Insulator with a Triangular Lattice}, Phys. Rev. Lett. {\bf 91}, 107001 (2003).
\bibitem{Kappa_BEDT_Specific_Heat} S. Yamashita, Y. Nakazawa, M. Oguni, Y. Oshima, H. Nojiri, Y. Shimizu, K. Miyagawa, and K. Kanoda, \textit{Thermodynamic properties of a spin$-1/2$ spin-liquid state in a $\kappa-$type organic salt}, Nat. Phys. {\bf 4}, 459 (2008).
\bibitem{Kappa_BEDT_Thermal_Conductivity} M. Yamashita, N. Nakata, Y. Kasahara, T. Sasaki, N. Yoneyama, N. Kobayashi, S. Fujimoto, T. Shibauchi, and Y. Matsuda, \textit{Thermal-transport measurements in a quantum spin-liquid state of the frustrated triangular magnet $\kappa-(BEDT-TTF)_{2}Cu_{2}(CN)_{3}$}, Nat. Phys. {\bf 5}, 44 (2009).
\bibitem{LSM_Theorem} E. Lieb, T. Schultz, and D. Mattis, \textit{Two soluble models of an antiferromagnetic chain}, Ann. Phys. (NY) {\bf 16}, 407 (1961).
\bibitem{LSM_Oshikawa} M. Oshikawa, \textit{Commensurability, Excitation Gap, and Topology in Quantum Many-Particle Systems on a Periodic Lattice}, Phys. Rev. Lett. {\bf 84}, 1535 (2000); M. Oshikawa, \textit{Topological Approach to Luttinger's Theorem and the Fermi Surface of a Kondo Lattice}, Phys. Rev. Lett. {\bf 84}, 3370 (2000).
\bibitem{LSM_Hastings} M. B. Hastings, \textit{Lieb-Schultz-Mattis in higher dimensions}, Phys. Rev. B {\bf 69}, 104431 (2004).
\bibitem{M_Cheng_LSM_Generalization} M. Cheng, \textit{Fermionic Lieb-Schultz-Mattis theorems and weak symmetry-protected phases}, Phys. Rev. B {\bf 99}, 075143 (2019).
\bibitem{Kanoda_MQCP_Transport} T. Furukawa, K. Miyagawa, H. Taniguchi, R. Kato, and K. Kanoda, \textit{Quantum criticality of Mott transition in organic materials}, Nat. Phys. {\bf 11}, 221 (2015).
\bibitem{DMFT_Mott_QCP} H. Terletska, J.Vucicevic, D. Tanaskovic, and V. Dobrosavljevic, \textit{Quantum Critical Transport near the Mott Transition}, Phys. Rev. Lett. {\bf 107}, 026401 (2011); J. Vucicevic, D. Tanaskovic, M. J. Rozenberg, and V. Dobrosavljevic, \textit{Bad-Metal Behavior Reveals Mott Quantum Criticality in Doped Hubbard Models}, Phys. Rev. Lett. {\bf 114}, 246402 (2015); V. Dobrosavljevic and D. Tanaskovic, \textit{Strongly Correlated Electrons in Two Dimensions}, edited by S. V. Kravchenko (Pan Stanford Publishing, 2017), pp. $1-46$.
\bibitem{Generic_Scale_Invariance} D. Belitz, T. R. Kirkpatrick, and T. Vojta, \textit{How generic scale invariance influences quantum and classical phase transitions}, Rev. Mod. Phys. {\bf 77}, 579 (2005).
\bibitem{Han_Cho_Kim_SLMQCP} Jae-Ho Han, Yong-Heum Cho, and Ki-Seok Kim, \textit{Spin-liquid Mott quantum criticality in two dimensions: Destabilization of a spinon Fermi-surface and emergence of one-dimensional spin dynamics}, Phys. Rev. B {\bf 95}, 235133 (2017).
\bibitem{Dangerously_Irrelevant_OP} H. v. Lohneysen, A. Rosch, M. Vojta, and P. Wolfle, \textit{Fermi-liquid instabilities at magnetic quantum phase transitions}, Rev. Mod. Phys. {\bf 79}, 1015 (2007).
\bibitem{U1SR_Original} S. Florens and A. Georges, \textit{Slave-rotor mean-field theories of strongly correlated systems and the Mott transition in finite dimensions}, Phys. Rev. B {\bf 70}, 035114 (2004).
\bibitem{U1SR_Kim} Ki-Seok Kim, \textit{Bandwidth-control versus doping-control Mott transition in the Hubbard model}, Phys. Rev. B {\bf 74}, 115122 (2006).
\bibitem{SU2SR_Kim} K.-S. Kim, \textit{SU(2) Gauge Theory of the Hubbard Model: Emergence of an Anomalous Metallic Phase near the Mott Critical Point}, Phys. Rev. Lett. 97, 136402 (2006); Ki-Seok Kim, \textit{How to control pairing fluctuations: SU(2) slave-rotor gauge theory of the Hubbard model}, Phys. Rev. B {\bf 75}, 245105 (2007); K.-S. Kim and M. D. Kim, \textit{Superconductivity from purely repulsive interactions in the strong-coupling approach: Application of SU(2) slave-rotor theory to the Hubbard model}, Phys. Rev. B {\bf 81}, 075121 (2010); M.-T. Tran and K.-S. Kim, \textit{Spin liquids in graphene}, Phys. Rev. B {\bf 83}, 125416 (2011).
\bibitem{anisotropy} A. Nakamura, Y. Yoshimoto, T. Kosugi, R. Arita, and M. Imada, \textit{Ab initio Derivation of Low-Energy Model for $\kappa-ET$ Type Organic Conductors}, J. Phys. Soc. Jpn. {\bf 78} 083710, (2009).
\bibitem{Lee_Lee_U1SL} S.-S. Lee and P. A. Lee, \textit{U(1) Gauge Theory of the Hubbard Model: Spin Liquid States and Possible Application to $\kappa-(BEDT-TTF)_{2}Cu_{2}(CN)_{3}$}, Phys. Rev. Lett. {\bf 95}, 036403 (2005).
\bibitem{LFL_RG} R. Shankar, \textit{Renormalization-group approach to interacting fermions}, Rev. Mod. Phys. {\bf 66}, 129 (1994).
\bibitem{Anomalous_Scaling_FS} S.-S. Lee, \textit{Stability of the U(1) spin liquid with a spinon Fermi surface in $2 + 1$ dimensions}, Phys. Rev. B {\bf 78}, 085129 (2008).
\bibitem{EFT_Large_N} S.-S. Lee, \textit{Low-energy effective theory of Fermi surface coupled with U(1) gauge field in $2 + 1$ dimensions}, Phys. Rev. B {\bf 80}, 165102 (2009).
\bibitem{Double_Patch_Construction_I} M. A. Metlitski and S. Sachdev, \textit{Quantum phase transitions of metals in two spatial dimensions. I. Ising-nematic order}, Phys. Rev. B {\bf 82}, 075127 (2010).
\bibitem{Double_Patch_Construction_II} M. A. Metlitski and S. Sachdev, \textit{Quantum phase transitions of metals in two spatial dimensions. II. Spin density wave order}, Phys. Rev. B {\bf 82}, 075128 (2010).
\bibitem{Nagaosa_Lee_SL} P. A. Lee and N. Nagaosa, \textit{Gauge theory of the normal state of high$-T_{c}$ superconductors}, Phys. Rev. B {\bf 46}, 5621 (1992).
\bibitem{Dimensional_Regularizarion_FS_I} D. Dalidovich and S.-S. Lee, \textit{Perturbative non-Fermi liquids from dimensional regularization}, Phys. Rev. B {\bf 88}, 245106 (2013).
\bibitem{Dimensional_Regularizarion_FS_II} S. Sur and S.-S. Lee, \textit{Quasilocal strange metal}, Phys. Rev. B {\bf 91}, 125136 (2015).
\bibitem{UV_IR_Mixing} I. Mandal and S.-S. Lee, \textit{Ultraviolet/infrared mixing in non-Fermi liquids}, Phys. Rev. B {\bf 92}, 035141 (2015).
\bibitem{Gauge_Symmetry_Correlation_Functions} Y.-B. Kim, A. Furusaki, X.-G. Wen, and P. A. Lee, \textit{Gauge-invariant response functions of fermions coupled to a gauge field}, Phys. Rev. B {\bf 50}, 17917 (1994).
\bibitem{RG_Textbook} M. E. Peskin and D. V. Schroeder, \textit{An Introduction to Quantum Field Theory} (Addison-Wesley Publishing Company, NewYork, 1995).
\bibitem{RG_Typical_Phi4} J. Zinn-Justin, \textit{Quantum Field Theory and Critical Phenomena}, (4th edition) (Oxford University Press, Oxford, 2002).
\bibitem{Luttinger_Liquid_Textbook} A. O. Gogolin, A. A. Nersesyan, and A. M. Tsvelik, \textit{Bosonization and Strongly Correlated Systems} (Cambridge University Press, New York, 2004).
\bibitem{IXY_Review} F. S. Nogueira and H. Kleinert, \textit{Field theoretic approaches to the superconducting phase transition, in Order, Disorder and Criticality: Advanced Problems of Phase Transition Theory}, edited by Y. Holovatch (World Scientific, Singapore, 2004), pp. 253.283.
\bibitem{SF_Mott_Transition_Review_I} V. Dobrosavljevic, N. Trivedi, and J. M. Valles Jr., \textit{Conductor Insulator Quantum Phase Transitions}, ISBN 9780199592593 (Oxford University Press, Oxford, 2012).
\bibitem{SF_Mott_Transition_Review_II} V. F. Gantmakher and V. T. Dolgopolov, \textit{Superconductor-insulator quantum phase transition}, Phys.-Usp. {\bf 53} 1 (2010).
\bibitem{Fracton_Review_I} R. M. Nandkishore and M. Hermele, \textit{Fractons}, Annu. Rev. Condens. Matter Phys. \textbf{10}, 295 (2019).
\bibitem{Fracton_Review_II} M. Pretko, X. Chen, and Y. You, \textit{Fracton phases of matter}, Int. J. Mod. Phys. A \textbf{35}, 2030003 (2020).
\bibitem{Xcube_fracton} S. Vijay, Jeongwan Haah, and L. Fu, \textit{Fracton topological order, generalized lattice gauge theory, and duality}, Phys. Rev. B \textbf{94}, 235157 (2016).
\bibitem{Haah_fracton} Jeongwan Haah, \textit{Local stabilizer codes in three dimensions without string logical operators}, Phys. Rev. A \textbf{83}, 042330 (2011).
\bibitem{Nematicity_Review} E. Fradkin, S. A. Kivelson, M. J. Lawler, J. P. Eisenstein, and A. P. Mackenzie, \textit{Nematic Fermi Fluids in Condensed Matter Physics}, Annu. Rev. Condens. Matter Phys. \textbf{1}, 153 (2010).
\bibitem{Coupled_Wire_Construction_Fracton} G. B. Halasz, T. H. Hsieh, and L. Balents, \textit{Fracton Topological Phases from Strongly Coupled Spin Chains}, Phys. Rev. Lett. \textbf{119}, 257202 (2017).

%
%\bibitem{HubbardModel} J. Hubbard, Proc. R. Soc. A, {\bf 276}, 1365 (1963).
%
\end{thebibliography}
\end{document}